\documentclass{aa}
\usepackage[varg]{txfonts}
\usepackage{natbib}
\begin{document}

\title{Simultaneous multi-frequency single pulse observations of pulsars}

\author{Arun Naidu\inst{1}
  \and Bhal Chandra Joshi\inst{1} 
   \and P.K Manoharan\inst{1,2}
   \and M.A KrishnaKumar\inst{1,2}} 

\offprints{Arun Naidu, \email{arun@ncra.tifr.res.in}, Bhal Chandra Joshi, \email{bcj@ncra.tifr.res.in}}

\institute{National Centre for Radio Astrophysics (Tata Institute for Fundamental
Research), P O Bag 3, Ganeshkhind, Pune - 411007 India
  \and Radio Astronomy Centre, NCRA-TIFR, Udhagamandalam (Ooty) 643001, India} 

\date{Accepted by A\&A}

\abstract {} {We report on simultaneous multi-frequency single pulse observations of a sample of pulsars with previously reported frequency dependent subpulse drift inferred from non-simultaneous and short observations. We aim to clarify if the frequency dependence is a result of multiple drift modes in these pulsars.}{We performed simultaneous observations at 326.5 MHz with the Ooty Radio Telescope and at 326, 610 and 1308 MHz with the Giant Meterwave Radio Telescope for a sample of 12 pulsars, where frequency dependent single pulse behaviour was reported. The single pulse sequences were analysed with fluctuation analysis, sensitive to both the average fluctuation properties (using longitude resolved fluctuation spectrum and two-dimensional fluctuation spectrum ) as well as temporal changes in these (using sliding two-dimensional fluctuation spectrum ) to establish concurrent changes in subpulse drifting over the multiple frequencies employed}{We report subpulse drifting in PSR J0934$-$5249 for the first time. We also report pulse nulling measurements in PSRs J0934$-$5249, B1508+55, J1822$-$2256, B1845$-$19 and J1901$-$0906 for the first time. Our measurements of subpulse drifting and pulse nulling for the rest of the pulsars are consistent with previously reported values. Contrary to previous belief, we find no evidence for a frequency dependent drift pattern in PSR B2016+28 implied by non-simultaneous observations by \cite{ohs77}. In PSRs B1237+25, J1822$-$2256, J1901$-$0906 and B2045$-$16, our longer and more sensitive observations reveal multiple drift rates with distinct $P_3$. We increase the sample of pulsars showing concurrent nulling across multiple frequencies by more than 100 percent, adding 4 more pulsars to this sample. Our results confirm and further strengthen the understanding that the subpulse drifting and pulse nulling are broadband consistent with previous studies \citep{Gajjar2014,Rankin86,wse07} and are closely tied to physics of polar gap.
}{} 


\maketitle

\section{Introduction}
\begin{figure}[h!]
\includegraphics[width=0.5\columnwidth]{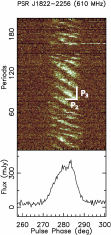}\centering
\caption{An example sequence of about 200 successive 
pulses of PSR J1822$-$2256 observed using the GMRT at 
610 MHz. The subpulses appear earlier with increasing 
pulse number and are arranged into so-called 
"drift bands". There are two distinct drift modes 
visible: a fast mode seen in the first 60 pulses 
followed by a slow mode. The two successive drift bands 
are vertically separated by $P_3$ periods and horizontally 
by $P_2$ degrees in phase as indicated for the slow mode. 
The slow drift mode is followed by a null between pulse 
number 144 and 176}
\label{figsubdr}
\end{figure}

While pulsars usually have a stable integrated profile 
[although some pulsars switch between two or three stable forms, 
or profile modes \citep{lyn71a}], their single pulses, 
consisting of multiple subpulses, often exhibit varying 
intensities and shapes on pulse to pulse basis. 
In some pulsars, these subpulses show a remarkable 
arrangement, where subpulses seem to be "marching" 
\citep{ssp+70} or drifting \citep{htt70} within the 
pulse window. This systematic progressive change in 
phase with the pulse number is called subpulse drifting 
(see Figure \ref{figsubdr}). Subpulse drifting is 
characterised by a periodicity along each longitude 
bin, $P_3$, and another periodicity along the pulse 
phase, $P_2$ (see Figure \ref{figsubdr}) with their 
ratio giving the drift rate. Some pulsars 
show a sharp significant drop in emission for several 
pulses called nulling, with the percentage of such 
pulses defined as the nulling fraction \citep{Ritchings1976}. 
Nulling sometimes also affects subpulse drifting 
\citep{la83,jv00}. In pulsars, such as PSR B0031$-$07 
and B2319+60, which show distinct drift modes, it 
has been shown that the different profile modes 
are associated with different drift rates 
\citep{wf81,vj97}. Pulse nulling itself can be 
considered as a form of profile mode-change.

Subpulse drifting is fairly common among pulsars 
\citep[][hereafter WES06 and WES07 respectively]{wes06,wse07}. 
Similarly, many pulsars are 
known to show nulling, with nulling known in more 
than 100 pulsars to date 
\citep{Wang2007,Biggs1992,Ritchings1976,Burke2012,gjk12,gjw14}. 
It is also believed that lack of emission for several 
hours to days in the recently discovered classes 
of pulsars, such as Rotating Radio Transients 
\citep{mll+06} and intermittent pulsars \citep{klo+06},  
can also be considered as an extreme form of nulling 
phenomenon. Lastly, profile mode-changes are known 
in several well studied pulsars \citep{Rankin86}, 
often associated with changes in the fluctuation 
properties of their single pulse emission 
\citep{Rankin86,wf81,vj97}.

Most previous studies concentrated on studying 
individual interesting pulsars for characterization 
of their drifting, nulling and mode-changing 
behaviour at a single observing frequency. 
The first systematic study of drifting at two frequencies 
were carried out at Westerbok Synthesis Radio Telescope 
(WSRT) almost a decade back (WES06,WES07). This  
study was useful in correlating the drifting 
behaviour of a large number of pulsars at two widely 
separated frequencies (21- and 92-cm waveband) and 
differences in the drift behaviour and drift 
rates were reported in several pulsars. These could be 
due to (a) differences in either intrinsic variability of 
emission process or emission geometry, leading to 
different drifting behaviour at two frequencies, 
or (b) presence of multiple drift modes and/or 
quiescent mode as seen in  
PSR B0943+10 \citep{so75} leading to a difference 
in the non-simultaneous short observations used for 
this study. This can only be resolved through 
a simultaneous multi-frequency study, which  
also helps in establishing generally a broadband 
nature of this phenomenon.

There are only a handful of simultaneous 
multi-frequency studies of single pulses available 
so far. Highly correlated pulse energy fluctuations 
were reported in a simultaneous single pulse study 
of two pulsars, PSRs B0329+54 and B1133+16 at 327 and 
2695 MHz \citep{bs78}. On the other hand, 
\cite{Bhatt2007} found that only half of nulls occur 
simultaneously at 325, 610, 1400 and 4850 MHz for 
PSR B1133+16. Three similar studies on nulling exist for 
PSRs B0031$-$07 and B0809+74 \citep{bar81,tmh75,dls+84}, 
whereas drifting in PSR B0031$-$07 was studied 
in one such study \citep{sms+07}. Recently, \cite{Gajjar2014}
mounted a major effort with three 
telescopes, the Giant Meterwave Radio Telescope (GMRT), 
the WSRT and the Effelsberg Telescope, covering 325 
to 4850 MHz, where strong evidence for concurrent 
nulls was found in PSRs B0031$-$07, B0809+74 and 
B2319+60. Thus, there is a need for a 
systematic sensitive simultaneous multi-frequency study 
of pulsars, which show drifting and pulse nulling, 
to enhance this sample. 

This paper presents results of a modest survey of 
simultaneous multi frequency observations of few 
selected pulsars, carried out using the 
Ooty Radio Telescope (ORT) and the GMRT  
utilizing frequencies from 326 MHz to 1300 MHz. 
Non simultaneous observations for three pulsars are 
also presented. We list the criteria for source selection 
in Section \ref{srcsel} followed by a description of 
observations and analysis procedure used in Sections 
\ref{obs} and \ref{anal} respectively. Investigation 
of subpulse modulations in each  observed 
pulsar are presented in Section \ref{res}. 
Discussion and conclusions follow in Sections \ref{disc} 
and \ref{conc} respectively.

\section{Source selection}
\label{srcsel}


We selected pulsars with an expected signal to noise ratio 
of more than 3 that can be observed using the ORT at 326.5 MHz 
and at least 7 antennas of a GMRT phased
array at 610 MHz. Signal to noise ratio was calculated 
using the radiometer equation, where the pulse width was 
adjusted to take into account the scatter broadening 
estimated using NE2001 model \citep{NE2001}. 
Additionally, the sky background was also taken into 
account using published all sky maps \citep{hss+82}
 in estimating the system 
equivalent flux density. The sample was further 
refined by restricting the declination range within 
$-53^{\circ}$ to $55^{\circ}$, which is the intersection of 
visible sky with both the ORT and the GMRT.  
After this selection, we have checked the 
literature for any previously available single pulse 
studies of the remaining pulsars in the list. We have 
selected those pulsars which have previously reported 
interesting single pulse behaviour, namely, prominent 
nulling or drifting as well as reported changes in the 
drift rates or multiple drift modes. 
Out of this list, we selected a subset of pulsars, where 
frequency dependent subpulse drift was reported in the past. 
PSR B1237+25 was chosen as a control pulsar to test the 
analysis pipeline developed by us. Only pulsars with 
no previously reported simultaneous multi-frequency 
single pulse study were selected. We were able to observe 
a total 12 pulsars during the time allocated for the 
survey (Table \ref{tabobs}).  


\section{Observations}
\label{obs}

All the observations were carried out using the ORT \citep{Swarup71}
and the GMRT \citep{sak+91}. The GMRT was used in a 
phased array mode with two sub-arrays consisting of 
about nine 45-m antennas, one at 610 MHz and the 
other at 1308 MHz bands, with a 33 MHz bandwidth. 
The phased array output for each of the two frequencies 
was recorded with 512 channels over the passband using 
the GMRT software baseband receiver with an effective 
sampling time of 1 ms along with a time stamp for the 
first recorded sample, derived from a GPS disciplined 
Rb frequency standard. The variations in the ionospheric 
and instrumental delays across the GMRT sub-arrays 
have a typical timescale of about 45 minutes at the 
observed frequencies. Hence, the observations were 
typically divided into two observing sessions, each 
of 45 minutes, interspersed with compensation for the 
instrumental delay drift to maintain phasing of the 
sub-array. The observations at the ORT were carried out 
using PONDER \citep{Naidu2015}. The observing band was 
centred at 326.5 MHz with a bandwidth of 16 MHz. 
The details of the observations are listed in 
Table \ref{tabobs}.


\begin{table*}
 \caption{Parameters of the observed pulsars and details of observations}
 \label{tabobs}
 \centering
 \begin{tabular}{ l c c c c c}
  \hline
  \hline
  Pulsar       & Period & Dispersion & Date of & Duration of  & Frequencies   \\
              &         & Measure & observations & observations & used \\
              &   (s)   & ($pc~cm^{-3}$) &         &  (minutes)     &  (MHz) \\
  \hline
  PSR J1901$-$0906  & 1.782 & 72.677 & 2013 May 05 & 90 & 610\\
  
  PSR J1901$-$0906  & 1.782 & 72.677 & 2013 May 19 & 90 & 325\\
  
  PSR J0934$-$5249 & 1.45 & 100.0 & 2013 Aug 08 & 30 & 325\\
  
  PSR J1822$-$2256 & 1.874 & 121.20 & 2014 Jan 31 & 60 & 610\\
      
  PSR B1237$+$25   & 1.382 & 9.2575  & 2014 Jun 18 & 30 & 326.5, 610\\
  
  PSR B1540$-$06  & 0.709 & 18.3774  & 2014 Jun 18 & 90 & 326.5, 610\\
  
  PSR B1844$-$04  & 0.598 & 141.979  & 2014 Jun 18 & 90 & 326.5, 610\\
  
  PSR B1508$+$55   & 0.740 & 19.6191 & 2015 May 05 & 90 & 326.5, 610, 1308\\
  
  PSR B1718$-$32   & 0.477 & 126.064 & 2015 May 05 & 90 & 326.5, 610, 1308\\
  
  PSR B1845$-$19   & 4.308 & 18.23 & 2015 May 05 & 90 & 326.5, 610, 1308\\
 
  PSR B2016$+$28   & 0.558 & 14.1977 & 2015 Sep 15 & 90 & 326.5, 610, 1308\\
  
  PSR B2043$-$04   & 1.547 & 35.80 & 2015 Sep 15 & 90 & 326.5, 610, 1308\\
  
  PSR  B2045$-$16  & 1.961 & 11.456  & 2015 Sep 15 & 90 & 326.5, 610, 1308\\
  \hline
 \end{tabular}
\end{table*}

\section{Analysis}
\label{anal}

\subsection{Pulse sequences}
\label{analpseq}

The data from both the observatories were converted into 
the standard format required for the 
SIGPROC\footnote{http://sigproc.sourceforge.net/} analysis 
package and dedispersed using the programs provided in the 
package. These were then folded to 1024 bins across the 
period using the ephemeris of these pulsars obtained with 
the TEMPO2 package to obtain a single pulse sequence (see 
Figure \ref{figsubdr}). The pulse sequences at different 
frequencies were aligned as follows. First, the pulse sequence 
for the longest data file, typically consisting of 2000 pulses, 
was averaged for each frequency to obtain an integrated 
profile, which was used to form a noise-free template after 
centring the pulse for the pulsar at that frequency. Samples 
from the beginning of each file were removed so that the 
pulse was centred in a period using the template for 
the corresponding frequency. Then, time stamps for single 
pulses were corrected by these offsets. These were converted to
solar system barycenter using 
TEMPO2\footnote{http://www.atnf.csiro.au/research/pulsar/tempo2/} 
\citep{hem06,ehm06} taking into account 
the delay at lower frequencies due to dispersion in 
the interstellar medium. Then, the pulses corresponding 
to identical time stamps at the solar system barycenter across all frequencies 
were extracted from the data to get pulse sequences 
aligned across frequencies. An example of such simultaneous 
multi-frequency single pulse sequences  
is given in Figure \ref{fig2016sp}. Plots of these pulse 
sequences for the complete sample are available as 
in the supplementary online material 
and in the electronic edition of the paper. 
The single pulse sequences were then 
visually examined to remove any single pulses with 
excessive radio frequency interference. 

For all the observation sessions at GMRT and few sessions 
at the ORT a suitable flux calibrator is observed for a short 
duration followed by a short observation of cold sky. The averaged
profile of the pulsar is calibrated with the appropriate scaling factor 
obtained from the ratio of averaged power on the flux calibrator to
averaged power of the cold sky (See 
bottom plot of Figure \ref{figsubdr}). For rest of the observations at the ORT, 
we used the average profile of the pulsar, obtained by 
adding all single pulses, an on-pulse and off-pulse 
window (including the phase, where the pulse was present 
and absent respectively) with equal number of bins were identified. 
A scaling factor was obtained by comparing the root-mean-square 
in the off-pulse window with the expected system equivalent 
flux density of the relevant telescope. The 
single pulse sequences were scaled with this factor and 
averaged to obtain a calibrated average profile. 
To estimate the nulling 
fraction, On-pulse and off-pulse energy sequences were formed 
by calculating the intensities in  all samples in the on-pulse 
and off-pulse windows. These were used to obtain on-pulse 
and off-pulse energy distributions. Nulling fractions were 
obtained from these distributions following the method 
described in \cite{gjk12}.

The pulse sequences were further analysed 
using the techniques discussed in following subsections.  

\subsection{Fluctuation Analysis}
\label{analfluc}

In bright pulsars, drifting subpulses, nulling and 
mode changing can be detected just by visual inspection 
of the  single pulse sequence. \cite{B70a,B70b} and \cite{B75} were 
the first to use the Fourier analysis techniques to 
characterise $P_3$. This often used technique is known as 
Longitude Resolved fluctuation spectrum (LRFS). While 
LRFS is useful to estimate $P_3$, it does not determine 
$P_2$, and hence the drift rate. \cite{es02} presented a 
modified technique called  Two-Dimensional Fluctuation 
Spectrum (2DFS) to simultaneously estimate $P_2$ and $P_3$. 
Both these methods describe drifting in an averaged sense 
and are insensitive to its temporal behaviour, important 
for our investigation. \cite{ssw09} developed a technique 
called Sliding two-dimensional Fluctuation Spectrum (S2DFS), 
based on the 2DFS, to provide information about the 
temporal changes that we are interested in. All three techniques 
are used by us and Figure \ref{figallplots} shows an example 
of application of these techniques. As part of this work, 
we developed a single pulse analysis code to implement these 
techniques and this code was used in the analysis of our 
observations. These techniques are 
explained briefly in the following sections.

\subsubsection{Longitude Resolved Fluctuation Spectrum}
\label{lrfs} 

\begin{figure*}
\includegraphics[width=\textwidth]{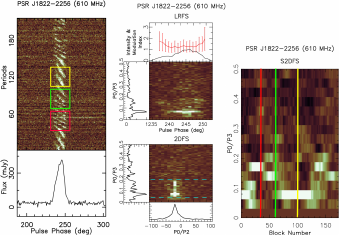}\centering
\caption{An example showing the results of analysis using the 
LRFS, 2DFS and S2DFS techniques for PSR J1822$-$2206. 
{\bf Left:} The single pulse sequence of PSR J1822$-$2206 
showing the drift bands with different drift rates and a 
null. {\bf Middle:} The LRFS and 2DFS plots of the single 
pulse sequence shown in the left. The top plot is  LRFS 
with the ordinate as $P_0$/$P_3$ and abscissa is the pulse 
phase. The top panel of LRFS is the integrated pulse profile 
along with the longitude-resolved 
modulation index (red line with the error bars). The bottom 
plot has the same ordinate as the top plot, but its 
abscissa is in units of $P_0$/$P_2$. The bottom panel in 2DFS 
shows the fluctuation frequency across a pulse integrated 
vertically between the indicated dashed lines around a feature. 
The left panels in the both LRFS and 2DFS plots are the spectra, 
integrated horizontally across the corresponding color plot. 
{\bf Right:} The $P_3$ S2DFS map made from the observation at 
610 MHz using the GMRT. The vertical axis is given in $P_0/P_3$. 
The horizontal axis is given in blocks, where a block 
corresponds average over N (typically 256) pulses. Periodic 
subpulse modulation is indicated by the "tracks" in this 
plot. The red, green and yellow zones indicate the blocks 
corresponding to zones marked with similar colour in the 
single pulse sequence shown in the leftmost plot.}
\label{figallplots}
\end{figure*}

The pulse sequence was divided into adjacent blocks of 
N pulses (typically N$\sim$256) and a Discrete Fourier 
Transform (DFT) was applied, at each pulse longitude bin 
in each block, to obtain LRFS for that block. The final 
spectrum is produced by averaging the LRFS of all blocks. 
Pulsars exhibiting periodically modulated subpulses will 
have a region, the so-called feature, of enhanced spectral 
power visible as a bright region in the LRFS. The middle 
top plot in Figure \ref{figallplots} shows the LRFS for 
the pulse sequence shown in the left plot of that figure. The 
frequency is given in cpp (cycles per period) and its 
inverse corresponds to the pattern periodicity, $P_3$, 
expressed in pulsar periods $P_0$. The position of the 
feature along the abscissa denotes the pulse longitude 
at which the modulation occurs. The subplot on top in 
the middle panel shows the integrated profile of 
PSR J1822$-$2206, normalised to the peak intensity,  
as solid line. The red line 
with the error bars in this plot is the longitude-resolved 
modulation index \citep[][]{es02}. The longitude-resolved 
modulation index is the measure of the amount by which the 
intensity varies from pulse to pulse for each pulse longitude.

\subsubsection{Two-Dimensional Fluctuation Spectrum.}
\label{2dfs}

The LRFS can be used only to estimate $P_3$. 2DFS is used to 
obtain both  $P_2$ and $P_3$. This method is similar to the 
calculation of the LRFS, but the DFT is applied twice. First 
DFT of the pulse sequence along the constant pulse longitude 
is recorded. Then, the DFT across each row of the complex LRFS 
is obtained. In Figure \ref{figallplots}, the 2DFS 
for observations of PSR J1822$-$2256 at 610 MHz is plotted 
below the LRFS plot. The vertical axis of the resulting 
spectra are the same as in the LRFS, but the horizontal 
axis now corresponds to the horizontal separation of the 
drift bands. If the drifting subpulses have a preferred 
drift direction, then a feature is seen offset from the 
vertical axis ($P_0/P_2  = $ 0). The 2DFS is vertically 
(between dashed lines) and horizontally integrated, 
resulting in the side and bottom panels. Estimates of 
$P_2$ and $P_3$, quoted in Tables \ref{nonsimul_stat} 
and \ref{simul_stat}, were calculated using  the 
centroid of a rectangular region in the 2DFS 
containing the feature. 

\subsubsection{Sliding two-dimensional Fluctuation Spectrum}
\label{s2dfs}

LRFS and 2DFS are very effective for detecting and 
analysing subpulse modulation. Integrating multiple 
fluctuation  spectra obtained from consecutive blocks 
of pulses increases the SNR of the resulting spectrum. 
However, this averaging does not reflect the 
temporal changes in drifting. To resolve the drift rate 
changes in the temporal domain, we have used the S2DFS 
technique developed by \cite{ssw09}. This method is an 
extension to the 2DFS. Here, 2DFS is computed for a block 
of a chosen number of pulses and collapsed over phase 
horizontally producing a longitude averaged fluctuation 
spectrum for the block (panel on left in 2DFS plot or 
LRFS plot in Figure \ref{figallplots}). The DFT window 
is then shifted by one pulse and the whole process is 
repeated, effectively sliding the window along the 
pulse sequence. This exercise of sliding the window 
and calculating the collapsed spectra, will result in 
N-M+1 curves, where N is the number of pulses in 
the pulse sequence  and M is the length of the DFT 
(block). These curves are arranged horizontally to 
form a map of collapsed fluctuation spectra called 
as S2DFS map. The right plot of Figure \ref{figallplots} 
shows an example of the S2DFS map, where one can easily 
see the "drift tracks" (regions of enhanced intensity 
visible as a bright region). The changes in periodic 
modulation with time, clearly visible in the color plot 
of this $P_3$ S2DFS map, reflect changes in drifting. 
We have used this analysis on pulse sequences, aligned 
across multiple frequencies, to study the simultaneity 
of changes in drifting across frequencies.

\begin{figure*}
\includegraphics[width=\textwidth]{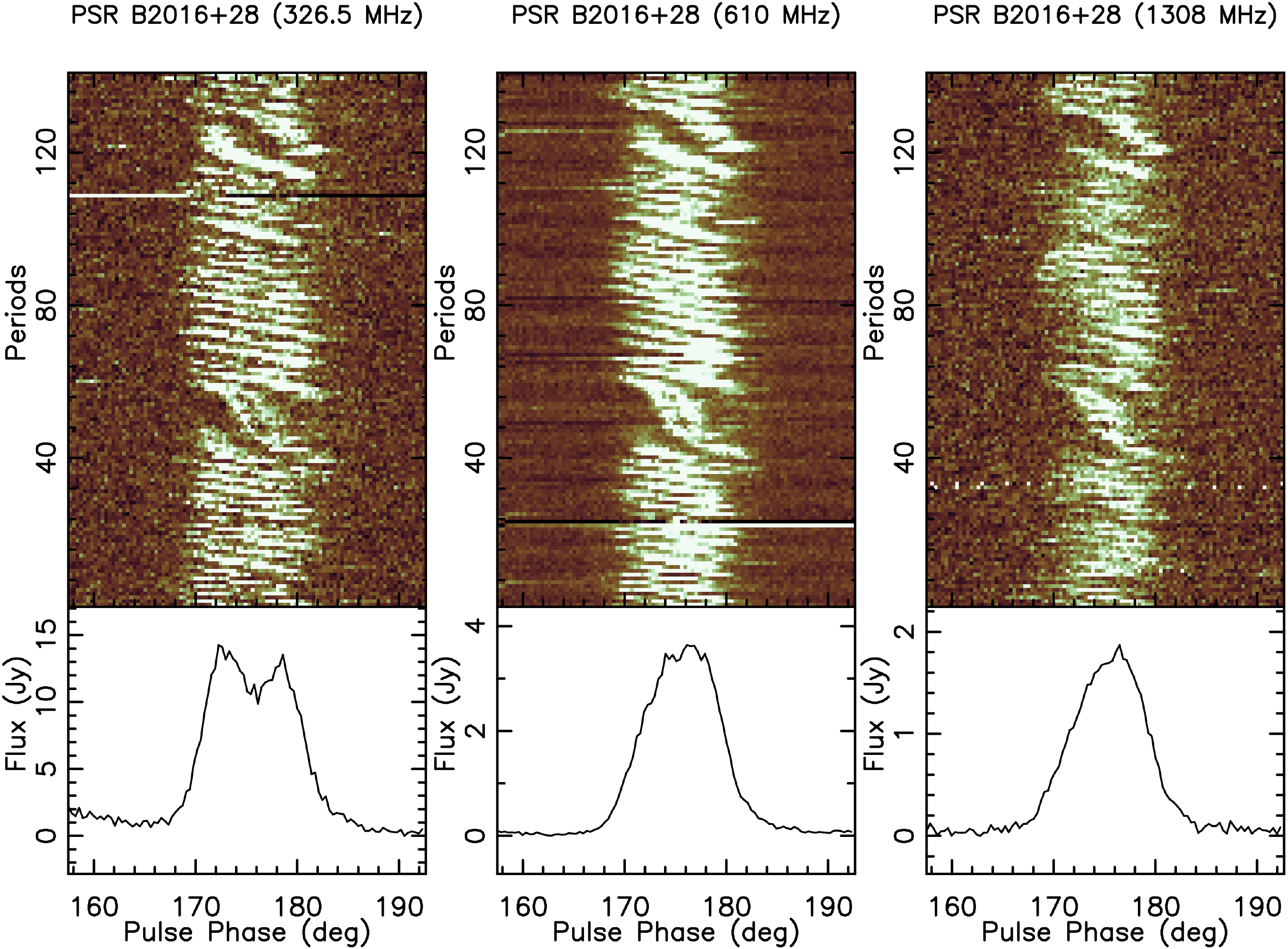}\centering
\caption{An example sequence of about 150 successive pulses 
of PSR B2016$+$28 observed simultaneously at three different 
frequencies (326.5 MHz with the ORT, and 610 and 1308 MHz with the 
GMRT).}
\label{fig2016sp}
\end{figure*}

The choice of  length of the DFT window is crucial 
for the resolution and sensitivity of these maps. For shorter 
Fourier transforms or a smaller DFT window, the S2DFS maps 
would lack the spectral resolution in $P_0/P_3$ required to 
resolve the changes in the drift rates. Conversely, a 
longer Fourier transform or larger DFT window would reduce 
the sensitivity to any short-lasting events, due to 
both the reduced SNR per spectral bin as well as a coarser temporal 
resolution. The duration of  drift modes varies from  pulsar 
to pulsar. Hence, the DFT length was selected by trial 
and error with block size ranging from 32 to 1024 pulses 
in logarithmic steps of 2. The optimum block size was 
selected by examining the corresponding pulse sequence 
plots and LRFS. In Figure \ref{figallplots}, we have 
used a window with 32 pulses as the duration of the 
fast drift mode is about 42 periods. 
The change in $P_3$ between the 
red, green and yellow windows is evident in 
Figure \ref{figallplots} marked on the pulse sequence 
in the plot on the left. It is important to note that 
during the transition from fast to slow drift the 
corresponding spectrum may be insensitive to both 
the modes.

\begin{figure*}
\includegraphics[width=0.53\textwidth]{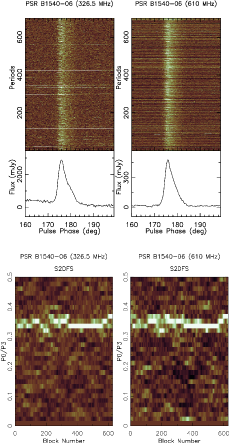}\centering
\caption{Example sequences of about 600 successive pulses of 
PSR B1540$-$06 observed at two different frequencies is presented 
in the top plot. The corresponding S2DFS plots using a window size 
of 64 pulses are shown in the lower plots.}
\label{1540_pseq_s2dfs}
\end{figure*}

The usefulness of S2DFS to investigate temporal subpulse 
drifting behaviour is apparent in Figure 
\ref{1540_pseq_s2dfs}. The single pulse sequences 
and corresponding S2DFS plots for PSR B1540$-$06 at 
two frequencies in simultaneous observations are 
shown in this Figure. The two S2DFS plots show 
changes in $P_3$ simultaneously at both the frequencies.  
Hence, this technique is very useful for our study 
concerning the broadband nature of the subpulse 
drifting for our sample of pulsars.


\section{Results}
\label{res}

The results of our analysis are presented in the 
Tables \ref{nonsimul_stat} and \ref{simul_stat}. 
The single pulse sequences,  
LRFS, 2DFS  and S2DFS plots for all 
pulsars along with the On-pulse energy sequences 
for 4 pulsars are available as an appendix   
in the supplementary online material. We highlight 
salient features of individual pulsars in this section.

\begin{table*}
 \caption{Modulation index and drift parameters for pulsars 
with independent multi-frequency observations. Pulsar name 
is given in the first column followed by its  
period, single pulse SNR, observation frequency, 
minimum modulation index, nulling fraction, $P_2$ and $P_3$.}
 \label{nonsimul_stat}
 \centering
 \begin{tabular}{ l c c c c c c c c}
  \hline
  \hline
  Pulsar & Period & Single   &  f$_{obs}$ & m  & NF & P$_2$      & P$_3$  \\
         &   (s)   & pulse &  (MHz)     &    & \% & $(^\circ)$ & (P$_0$) \\
         &        & SNR    &           &    &          &       &          \\

\hline
J0934$-$5249 & 1.45 & 2.1 & 325 & $0.96\pm0.07$ & $5\pm3$& $-7^{+2}_{-3}$   &$4.0\pm0.1$ \\
\hline
J1822$-$2256 & 1.874& 1.8 & 610 & $1.2\pm0.5$& $10\pm2$& $-13^{+1}_{-2}$   &$17\pm1$ \\
&&&&&&                                                    $-14^{+1}_{-2}$   &$7.5\pm0.4$ \\
&&&&&&                                                    $-13^{+1}_{-2}$   &$6.3\pm0.3$ \\
\hline
J1901$-$0906 &1.782 & 1.3 &325 & $1.4\pm0.2$ & $29\pm4$ & $-9^{+1}_{-2}$ &$3.0\pm0.3$ \\     
& & & & & &                                           $-25^{+6}_{-13}$ & $5.4\pm0.1$ \\
& & & & & &                                           $-17^{+2}_{-3}$ & $7.3\pm0.2$ \\ 
 &  &                 1.6    &610 & $1.2\pm0.1$ &$30\pm1$& $-8^{+1}_{-1}$   &$3.1\pm0.1$ \\     
& & & & & &                                           $-26^{+7}_{-13}$ & $5.1\pm0.3$ \\
& & & & & &                                           $-12^{+1}_{-2}$ & $7.6\pm0.6$ \\  \hline
 \end{tabular}
\end{table*}

\subsection{Non-simultaneous observations}
\label{resnonsim}

\noindent
{\bf PSR J0934$-$5249}: This pulsar was observed at a single 
frequency (325 MHz) using the GMRT. Its single pulse sequence 
shows clear coherent drifting with some short nulls 
(Figure \ref{A:0934_sp}). 2DFS (see Figure \ref{A:0934_2dfs}) 
plot also shows a clear feature confirming 
that  this pulsar is a coherent drifter. The number of 
nulls were small for us to study any null induced 
drift rate changes in this pulsar. Estimates for 
drift and nulling parameters are being reported for 
the first time (Table \ref{nonsimul_stat}) in this pulsar.

\noindent
{\bf PSR J1822$-$2256}: This pulsar was observed at a 
single frequency (610 MHz) using the GMRT. It shows
regular drifting with changes in drift rates as is 
evident in Figure \ref{figallplots}. The 2DFS plot in 
this figure shows a clear feature. However, the drift 
rate changes seen in the single pulse sequence are 
not clearly visible in the 2DFS plot, probably due to 
dominance of the slower drift mode in the selected 
pulse sequence. Here, S2DFS analysis is clearly very 
useful as it not only brings out two almost harmonically 
related drift modes, but also temporal time scale for 
these drift modes. Our results confirm the 
two drift rates seen by WES06 at 21-cm. Later studies 
by WES07 and \cite{bmm+16} detected only one drift 
rate ($P_3 \sim 17 P_0$). While 
our observations were at 610 MHz only, we detect 
both modes in our observations. The previous non-detection 
at lower frequency is probably due to short observations, 
where only one mode might have been present. Hence, our  
result along with the presence of multiple drift modes 
implies a broadband drift behaviour. 
The single pulse sequence shows 
regular nulls leading confirming reported pulse 
nulling by \cite{Burke2012}. The nulling fraction 
for this pulsar is being reported for the first time and  
is $10\pm2 \%$. The single pulse properties of this pulsar 
are very similar to PSR B0031$-$07 \citep{vj97} due to 
presence of nulling and harmonically related drift modes. 

\noindent
{\bf PSR J1901$-$0906}: This pulsar was observed at two 
different frequencies (325 MHz and 610 MHz) 
independently using the GMRT. The integrated profile 
has two widely separated components (See Figure 
\ref{A:1901_sp}). The drifting is clearly visible in the 
pulse sequence as is pulse nulling. WES06 first reported 
drifting in this pulsar at 21-cm and concluded that 
it shows drifting with two distinct values of $P_3$ (3 and 7 $P_0$). 
They detected the former mode only in the trailing 
component, while the latter was seen in both  
components. WES07 also detected drifting at 92-cm with 
two different $P_3$ (3 and 5 $P_0$) with the latter value 
being different from that seen at 21-cm. These authors 
concluded that this pulsar shows differences in sub-pulse 
drift not only in its two components, but also between 
the two frequencies. Our longer and higher sensitivity 
observations suggest that the pulsar exhibits {\it {\bf three}} 
distinct drift modes in {\it {\bf both}} the components 
(see Table \ref{nonsimul_stat}) at {\it {\bf both}} 325 
and 610 MHz. The fast drift mode is more prominent in 
the trailing component, while the slow modes are more 
prominent in the leading component (See 
Figure \ref{A:1901_2dfs}). This is probably one of the reasons 
for difference in drift properties of the two components 
seen by WES06. The different measured P$_3$ values in the 
two components confirms that this pulsar is a drift 
mode changer. Interestingly, the two components do not 
exhibit subpulse drift with identical $P_3$ as can be 
seen in S2DFS plots of the two components (Figure 
\ref{A:1901_s2dfs}). 
The frequency dependent and component dependent drift 
reported by WES06 and WES07 most likely arises due to this 
complex sub-pulse behaviour seen in our observations. Our 
observations suggest a broadband drifting behaviour in this 
pulsar. 
J1901$-$0906 also shows several prominent nulls. We are 
reporting nulling fraction for this pulsar for the first time 
(Table \ref{nonsimul_stat}). It may be noted that this is a unique and 
interesting pulsar exhibiting not only multiple 
distinct drift modes, similar to PSR B0031$-$07, but also 
exhibiting different drift modes in two widely separated 
components. Future long simultaneous multi-frequency 
observations of this pulsar are motivated.

\subsection{Simultaneous observations}
\label{ressim}

\noindent
{\bf PSR B1237$+$ 25}: This pulsar has a multi-component 
profile and was observed simultaneously at 326.5 MHz and 610 MHz 
using the ORT and the GMRT respectively. Single pulse sequences 
at both frequencies are shown in Figure \ref{A:1237_sp} 
indicating a correlated single pulse behaviour 
including nulls at both frequencies. On-pulse 
energy sequences, shown in Figure \ref{A:1237_ep}, 
suggest that the pulsar nulls simultaneously at both  
frequencies. Nulling fraction estimated from our observations 
(Table \ref{simul_stat}) are consistent with that 
reported previously \citep{Ritchings1976}. WES06 reported 
a detection of fast mode ($P_3 \sim 2.8 P_0$) in 
all components at 21-cm and in three components 
at 92-cm. We have detected a similar mode at both 
326.5  and 610 MHz in the leading and trailing 
components (Figure \ref{A:1237_2dfs}). We also 
detect a slow mode ($P_3 \sim 28\pm8~P_0$) in the 
central component at 326.5 MHz consistent with that 
reported by \cite{md14} (see Figure \ref{A:1237_2dfs_comp2}). 
Simultaneous 610 MHz  data show the same feature 
but with lower intensity. Thus, our observations are 
consistent with previous studies of drifting in this 
pulsar. In addition, S2DFS analysis shows that the 
temporal behaviour of the drifting subpulses is 
similar at both the frequencies (See Figure 
\ref{A:1237_s2dfs}). 

\noindent
{\bf PSR B1508$+$55}: Observations were carried out at 
three frequencies. The single pulse SNR was low at 326.5 
and 1308  MHz, whereas observations at 610 MHz show strong 
single pulses (Figure \ref{A:1508_sp}). 2DFS plots show a 
broad low frequency feature. Pulsar shows clear  nulls in 
single pulse sequences, which appear to be correlated 
across 326.5 and 610 MHz (Figure \ref{A:1508_ep}). We report 
pulse nulling for the first time in this pulsar and nulling fraction
is estimated to be $7\pm2\%$.

\noindent
{\bf PSR B1540$-$06}: The simultaneous observations of 
this pulsar were carried out at 610 and 326.5 MHz 
respectively. There is strong feature at $3.0\pm0.2~P_0$
in the 2DFS plot at both frequencies (Figure 
\ref{A:1540_2dfs}). Our measurement is consistent 
with previous studies (WES07). The S2DFS plots show that $P_3$  
varies a bit, but the changes in drift are simultaneous 
(Figure \ref{A:1540_s2dfs}). There appear to be some 
nulls, but SNR at 326.5 MHz was not sufficient to 
detect such nulls. Nulls are more discernible 
at 610 MHz and we estimate the nulling fraction to be about 6 percent 
(Figure \ref{A:1540_sp}).

\noindent
{\bf PSR B1718$-$32}: Observations of this pulsar were 
carried out at three different frequencies (326.5, 
610 and 1308 MHz). The single pulse SNR was low at 
326.5 MHz and 1308 MHz, but strong single pulses were 
seen at 610 MHz (Figure \ref{A:1718_sp}). The 2DFS plot 
(Figure \ref{A:1718_2dfs}) shows a bright feature at 
about $22\pm3~P_0$. Visual examination of single 
pulse sequence reveals that this feature is due to 
amplitude modulation of the leading component and 
does not appear like subpulse drifting. This modulation 
feature is being reported for the first time. No significant nulling is 
detected in this pulsar.

\noindent
{\bf PSR B1844$-$04}: The single pulse SNR for this  
pulsar at the ORT was low and only single pulses 
from the GMRT observations were useful for 
further analysis (Figure \ref{A:1844_sp}). The LRFS 
shows a broad low frequency feature at $11\pm2~P_0$ (see 
Figure \ref{A:1844_2dfs}). WES06 detected this feature 
at 21-cm waveband but not at 92-cm waveband. The profile at 326.5 MHz 
is scatter-broadened, which masks subpulse drifting 
and this could be the reason for non-detection of a drift 
feature in our as well as past observations.


\noindent
{\bf PSR B1845$-$19}: No significant drifting is observed 
at any of the three frequencies. The pulsar nulls 
frequently and nulls are simultaneous at 326.5 and 610 MHz 
(Figures \ref{A:1845_sp} and \ref{A:1845_ep}). Pulse 
nulling is being reported for the first time in this 
pulsar with nulling fraction estimated to be $27\pm6 \%$.

\noindent
{\bf PSR B2016$+$28}: This pulsar shows prominent drift 
bands at all three frequencies. The drift pattern is 
variable, but correlated across the three frequencies 
(Figure \ref{fig2016sp}). This is in contrast with results 
reported by \cite{ohs77} suggesting a frequency dependent 
subpulse drift. Indeed, a feature detected in 2DFS plot 
at 1308 MHz is not apparent at 326.5 MHz 
(Figure \ref{A:2016_2dfs}), consistent with a similar result 
reported by WES06 and WES07. However, a weak feature is 
present at 610 MHz (and possibly at 326.5 MHz) in the 
2DFS plot for the leading component of the integrated 
profile for this pulsar (Figure \ref{A:2016_2dfs_comp1}). 
Moreover, there are few sections of data, where this component 
is detected at all three frequencies as is evident in S2DFS plot 
for one such section (Figure \ref{A:2016_s2dfs_1}). Lastly, 
the temporal changes in the S2DFS plot (Figure \ref{A:2016_s2dfs}) 
indicate that the drift pattern changes simultaneously at all 
frequencies. As the drift rate varies from band 
to band resulting in the broad features in the LRFS, analysis 
of short and/or non-simultaneous observations can mimic a frequency 
dependence, which is not borne out by our longer simultaneous 
observations. Thus, we conclude that drifting is independent of 
frequency of observations. The pulsar does not show any significant 
nulling.

\noindent
{\bf PSR B2043$-$04}: The single pulse SNR of this pulsar 
at 1308 MHz was low during our observations (Figure 
\ref{A:2043_sp}), but 326.5 and 610 MHz data show a strong 
feature at $2.7\pm0.1~P_0$ in our 2DFS analysis (Figure 
\ref{A:2043_2dfs}). Although single pulses are not visible 
at 1308 MHz, this feature is also present in 2DFS at 1308 MHz. 
There is also a very weak feature at $3.7~P_0$, seen in 
2DFS plots (see Figures \ref{A:2043_2dfs} and 
\ref{A:2043_2dfs_comp1}). $P_3$ varies sometimes with 
these changes occurring simultaneously at 326.5 and 610 
MHz (Figure \ref{A:2043_s2dfs}). No significant  nulls are 
observed in this pulsar.

\noindent
{\bf PSR B2045$-$16}: This pulsar has a multiple component 
profile. A variety of single pulse behaviour along with 
prominent nulls is seen in the single pulse plots (Figure 
\ref{A:2045_sp}). A broad feature with $P_3$ values 
between 2  and 3 $P_0$ were reported in the outermost 
components in some previous studies \citep{ohs77,nuk+82}, 
whereas drifting with $P_3 = $ 3.2 $P_0$ was reported 
ONLY in the trailing component at 21-cm waveband (WES06). 
In contrast, WES07 reported drifting in three components 
at 92-cm waveband with $P_3$ varying between 2.7 to 3 
$P_0$ as well as a low frequency feature with $P_3 = $ 32 
$P_0$. Our 2DFS analysis shows a strong 
feature at $3.2~P_0$ in the leading and trailing 
components (See Figure \ref{A:2045_2dfs}). These features 
are also evident in S2DFS plots for 326.5 and 610 MHz 
(Figure \ref{A:2045_s2dfs}). These plots indicate that 
the fluctuation frequency varies between $3.57~P_0$ to $2.5~P_0$ 
in all components and the changes in $P_3$ occur simultaneously 
across all frequencies, including 1308 MHz. where WES06 reported 
drifting only in the trailing component. Thus, drifting 
behaviour appears to be broadband in this pulsar and  
any differences reported in the past can be attributed 
to drift modes in this pulsar and non-simultaneity of 
observations. The pulsar exhibits nulling and its nulling fraction is  
tabulated in the Table \ref{simul_stat}. The nulls seem 
to be broadband (Figure \ref{A:2045_ep}) and the nulling 
fraction is consistent across all frequencies.

\begin{table*}[h]

\caption{Modulation index and drift parameters for pulsars with simultaneous multi-frequency observations. Pulsar name is given in the first column followed by corresponding period, single pulse SNR, observation frequency, minimum modulation index, nulling fraction, $P_2$ and $P_3$.}
 \label{simul_stat}
 \centering
 \begin{tabular}{ l c c c c c c c c}
  \hline
  \hline
  Pulsar & Period & Single Pulse   &  f$_{obs}$ & m  & NF     & P$_2$ & P$_3$  \\
         &   (s)     & SNR          &  (MHz)     &    & (\%)&  ($^\circ$) & (P$_0$)   \\
\hline
B1237$+$25& 1.382 & 1.7 & 325 & $1.1\pm0.2$ & $7\pm3$ & $-35^{+10}_{-12}$ &$2.8\pm0.8$ \\         
& & & & & &                                     $-15^{+11}_{-15}$   &$2.8\pm0.2$ \\
& & & & & &                                     $-12^{+2}_{-19}$   &$28\pm8$ \\
      &          & 5.0 & 610 & $0.57\pm0.06$ & $4\pm1$ & $-31^{+2}_{-2}$ & $2.8\pm0.1$ \\
& & & & & &                                      $-13^{+8}_{-23}$ & $2.9\pm0.2$ \\
& & & & & &                                     $-13^{+12}_{-27}$   &$29\pm8$ \\
\hline
B1508$+$55 & 0.740 & 0.4 & 325 &$1.3\pm0.1$&   &    &        \\
&                   & 3.5 & 610 & $0.75\pm0.06$ & $7\pm2$ &    &   \\
&                   &  0.1 & 1308 &  &   &    &        \\
\hline
B1540$-$06  & 0.709 & 2 & 325 & $1.0\pm0.1$ & $2\pm1$&  $15^{+30}_{-61}$ & $3.0\pm0.2$ \\
&                   & 1.4 & 610 & $0.72\pm0.06$& $4\pm2$ & $6^{+10}_{-30}$ & $3.0\pm0.1$ \\
\hline
B1718$-$32   & 0.477 & 0.06 & 325 & - &   &    &    &    \\
&                   & 1.4 & 610 &  $0.69\pm0.04$ & $1\pm1$ & $-8^{+3}_{-12}$ & $22\pm3$ \\
&                   &  0.1 & 1308 & - &   &    &        \\
\hline
B1844$-$04  & 0.598 & 1.7 & 610 & $0.7\pm0.3$ & $3\pm1$ & $-10^{+42}_{-21}$   &$11\pm2$ \\ 
     &            & 0.1 & 325 & & - &                &        \\          
\hline
B1845$-$19   & 4.308& 0.3 & 325 & $1.7\pm0.6$ & $27\pm6$ &    &        \\
&                   & 3.6 & 610 &  $1.2\pm0.3$ & $19\pm4$ &    &   \\
&                   &  0.1 & 1308 & -  &   &    &        \\
\hline
B2016$+$28   & 0.558& 4 & 325 & $1\pm0.5$ & $1\pm2$  & $-8^{+1}_{-1}$   &        \\
&                   & 9 & 610 &  $0.7\pm0.3$ & $2\pm2$   & $-8^{+1}_{-2}$   &   \\
&                &  1.5 & 1308 & $0.6\pm0.2$ & $1\pm3$& $-9^{+4}_{-7}$   & $21\pm4$  \\
\hline
B2043$-$04 & 1.547 & 1.4 & 325 & $1.1\pm0.2$ &  & $3^{+15}_{-7}$ & $2.7\pm0.1$  \\
&                   & 2 & 610 &  $0.9\pm0.1$ &  & $0^{+18}_{-1}$ & $2.7\pm0.1$\\
&                   &  0.2 & 1308 &   &   &     &      \\
\hline
B2045$-$16 & 1.961& 1.6 & 325 & $1.1\pm0.2$ &$14\pm3$ & $-35^{+9}_{-15}$ & $3.2\pm0.2$   \\
&   &      &       &        &                & $-15^{+3}_{-3}$ & $3.5\pm0.5$     \\
&                   & 2.8 & 610 &  $1.2\pm0.2$ & $17\pm6$& $-35^{+7}_{-12}$& $3.2\pm0.2$ \\
&   &      &       &        &                      & $-16^{+2}_{-3}$ & $3.5\pm0.6$   \\
&                 &  0.9 & 1308 & $1.2\pm0.2$ & $22\pm5$& $-29^{+10}_{-47}$& $3.3\pm0.3$ \\
&   &      &       &        &                & $-19^{+5}_{-13}$   & $3.4\pm0.5$     \\
\hline
\end{tabular}
\end{table*}

\section{Discussion}
\label{disc}

The results from our simultaneous multi-frequency single 
pulse observations of nine pulsars with subpulse drifting 
or nulling are presented in this paper. We also report on 
single frequency observations for three pulsars. We report subpulse 
drifting in PSR J0934$-$5249 for the first time. We also report
pulse nulling measurements in PSRs J0934$-$5249, B1508+55, J1822$-$2256, 
B1845$-$19 and J1901$-$0906  for the first 
time. Our measurements of subpulse drifting and pulse nulling 
for the rest of the pulsars are consistent with previously 
reported values.

Most of the pulsars in our sample were observed at two or more frequencies 
{\it {\bf simultaneously}}. We have made an attempt to 
understand the fluctuation properties of these pulsars 
by examining single pulse sequences, LRFS, 2DFS and S2DFS 
analysis. To examine the temporal changes in drift 
pattern, we have used S2DFS method. The simultaneous 
temporal changes are investigated across the frequencies 
with both visual as well as S2DFS analysis to check for 
any frequency dependent behaviour. 

Our results confirm and further strengthen the conclusions 
drawn by WES06 and WES07, 
where these authors state that subpulse drifting is broadband 
in general. Our sample consisted of pulsars with reported 
differences in $P_3$ at different frequencies by these authors 
and by other past studies. For example, non-simultaneous short 
observations by WES06 and WES07 suggested different $P_3$ values 
at 21-cm and 92-cm 
for PSRs J1822$-$2256, J1901$-$0906, B1844$-$04, B2016+28 and 
B2045$-$16. Similarly, drifting was seen in only one component 
of the integrated profile in PSRs J1901$-$0906, B2016+28 and 
B2045$-$16 at one or both  frequencies in their study. Frequency 
dependent subpulse drifting was suggested in at least two past 
studies \citep[][WES06,WES07]{ohs77}. As mentioned before, 
this could be due to (a) dependence of subpulse drifting 
mechanism on emission height (and therefore observing frequency 
by virtue of radius-to-frequency mapping), (b) geometric origin 
manifested in profile evolution or (c) presence of drift modes 
or variation in drift rate leading to an apparent difference 
due to short duration observations and non-simultaneous nature 
of these studies. In our work, we have examined the drift behaviour 
with longer (typically 90 minutes) simultaneous multi-frequency 
observations to distinguish between these three possibilities. 
Contrary to previous belief, we find no evidence for a frequency 
dependent drift pattern in PSR B2016+28 implied by 
non-simultaneous observations by \cite{ohs77}. In PSR 
B1237+25, J1822$-$2256, J1901$-$0906 and B2045$-$16, 
our longer and more sensitive 
observations reveal multiple drift rates with distinct $P_3$, 
consistent with the values reported previously using observations 
where probably only a given mode was present. Additionally, our 
S2DFS analysis of  pulse sequences aligned across frequencies 
show changes in $P_3$ occurring at the same time across frequencies 
for these pulsars. This is also true for other pulsars,  
except PSR B1844$-$04, where scatter-broadening 
masks drifting at 326.5 MHz. Thus, we conclude that subpulse drift 
is broadband even in these pulsars and multiple drift modes can 
give appearance of a frequency dependence if the (a) observation 
duration is smaller than the time scales required to sample all 
modes, or (b) a given drift mode is rare. The implied broadband nature 
also suggests that geometry of pulsar emission including variations 
with emission heights are unlikely to affect the drift periodicities. 

Some of the pulsars in our sample also exhibit pulse nulling. 
The single pulse sequences were visually examined and 
nulling seemed to be simultaneous across all the observed 
frequencies. The nulling fractions at different frequencies 
are consistent. Thus, pulse nulling appears to be broadband 
in these pulsars. A recent multi-frequency study of 3 pulsars 
with long observations has reported broadband nulling in those 
pulsars \citep{Gajjar2014}. We add 4 more pulsars to this 
list increasing the sample of such pulsars by more than 100 
percent.

Establishing a broadband nature of these phenomena is 
important for the following reasons. Profile evolution 
with frequency implies that the frequency of choice 
for studies of drifting can be different for 
different pulsars. Once, it is broadly established 
that drifting (and nulling behaviour) is concurrent 
across the observing frequency, an appropriate frequency 
can be chosen for more sensitive observations. 
Moreover, deeper studies can then be taken to 
characterize changes in drift behaviour, 
which may be related to profile mode-changes. 
Recently, it has been shown that off state in 
intermittent pulsars as well as pulsars with profile 
mode-changes are accompanied by changes in spin-sown 
rate \citep{lhk+10}. Long follow up observations with 
appropriately chosen frequency will therefore be 
very useful in understanding switching behaviour of 
magnetosphere, which has been invoked to explain 
the spin-down changes mentioned above \citep{Timokhin2010}.
With multi-beam capability in upcoming large collecting 
area telescopes, such as Square Kilometer Array (SKA), 
such long term observations are possible and our study 
allows careful choice of frequency band in SKA for this 
purpose.  

The lack of corotation of localized ``sparks'' 
in the polar gap, defined by the open field lines, 
was invoked by \cite{RS75} to explain the subpulse 
drift. 
In this model, the relativistic outflow of electron-positron from 
these “sparks” produces the subpulse associated radio emission, 
higher up in the magnetosphere. The local plasma frequency 
and magnetic field at a given height determines the frequency 
of radio emission emitted at that height, which 
is also refered to ``radius-to-frequency mapping'' \citep{cor78}.  
As the ``sparks'' lag behind the rotation of the star, 
the associated radio emission also progressively 
changes its phase within the period, which is seen as 
subpulse drift, with the subpulses associated with each 
``spark''. In this model, the drift is likely to 
be similar at different frequencies as it is determined 
solely by the motion of ``sparks'', which is governed 
by the gap potential and magnetic field in the polar gap. 
Hence, subpulse drift is broadband in this model  
in contrast with models invoking dependence on viewing 
geometry.

Finally, in \cite{RS75} model, the drift rate of subpulses provides a 
fundamental probe of the polar cap physics 
as well as the changes in the state of 
magnetosphere. If profile mode-changes 
(including nulls) are related to drifting, 
as established in PSR B0031$-$07 \citep{vj97} and 
B2319+60 \citep{wf81}, a study of these 
phenomena over multiple frequencies can provide 
useful constraints on magnetospheric physics. 
Our study points out three more candidates for such studies. 
These are PSR J1822$-$2256, J1901$-$0906 and B2045$-$16, 
where longer observations, similar to us are motivated. Upcoming 
telescopes like SKA can enhance this sample substantially 
apart from providing higher sensitivity pulse sequences.

\section{Conclusions}
\label{conc}

The observations presented in this paper  provide further 
confirmation that the subpulse drift and pulse nulling 
are predominantly broadband consistent with previous studies 
\citep[][WES07]{Gajjar2014,Rankin86}. While this is expected 
from models such as \cite{RS75}, we have dealt with 
some of previously reported exceptions 
in this paper and find that any frequency dependent behaviour 
can be attributed to multiple drift modes. Our results, thus 
suggest that the origin of drifting and pulse nulling is closely 
tied to physics of polar gap. 

\section{Acknowledgments}

We thank the staff of the Ooty Radio Telescope and the Giant Meterwave
Radio Telescope for making these observations possible. Both these
telescopes are operated by National Centre for Radio Astrophysics (TIFR).
This work made use of PONDER backend, built with TIFR XII plan grants
12P0714 and 12P0716. We like to thank the anominous referee for his useful
comments and suggestions. AN like to thank Dave Green and Yogesh Maan for providing
the color scheme used for the plots. BCJ, PKM and MAK acknowledge support for this work
from DST-SERB grant EMR/2015/000515.

\bibliographystyle{aa}
\renewcommand{\bibname}{References} 

\bibliography{references} 
 
\newpage
\begin{appendix}
\section{Analysis plots}
 All the relevant plots from our analysis is given in this appendix. Plots are arranged according to the pulsar name. A brief description of techniques and the plots is given below.
  
\begin{enumerate}
\item {\bf Single Pulse Sequence:} These plots are used for visual examination of the single pulse sequences for drifting and nulling behaviour. The color plot is the stack of single pulses and the bottom panel is the integrated profile with abscissa is the pulse phase in degrees and the ordinate is the pulse flux density in milli Jansky.

\item {\bf Longitude resolved fluctuation spectra (LRFS):} This plot is used to estimate pattern separation (P$_3$). The color plot is the LRFS with the ordinate as $P_0$/$P_3$ (P$_0$ is the period of the pulsar) and abscissa is the pulse phase. The top panel of LRFS is the integrated pulse profile along with the longitude-resolved modulation index (red line with the error bars). The plot on left is the longitude averaged fluctuation spectra.

\item {\bf 2-Dimentional Fluctuation Spectra (2DFS):} The subpulse separation (P$_2$) is estimated using the 2DFS. The color plot is the 2DFS with the ordinate as $P_0$/$P_3$ and abscissa is the in units of $P_0$/$P_2$. The bottom panel in 2DFS shows the fluctuation frequency across a pulse integrated vertically between the indicated dashed lines around a feature.

\item {\bf Sliding 2-Dimentional Fluctuation Spectra (S2DFS):} This technique is used to estimate the temporal variation of the P$_3$. This plot is obtained by stacking the longitudinal averaged 2DFS spectra for pulses in the short imaginary window sliding along the pulse sequence. The ordinate of the color plot is $P_0$/$P_3$ and abscissa is corresponding block number.

\item {\bf On-pulse sequence:} These plots can be used to visualize the pulse energy variation across the frequencies. The ordinate is the significance of the on-pulse energy in units of standard deviation of the off-pulse energy and the abscissa is the pulse number. The black line in all these plots in the off-pulse energy.  
 
\end{enumerate}

\begin{figure*}[!tbp]
  \centering
  \begin{minipage}[b]{0.2\textwidth}
    \includegraphics[width=\textwidth]{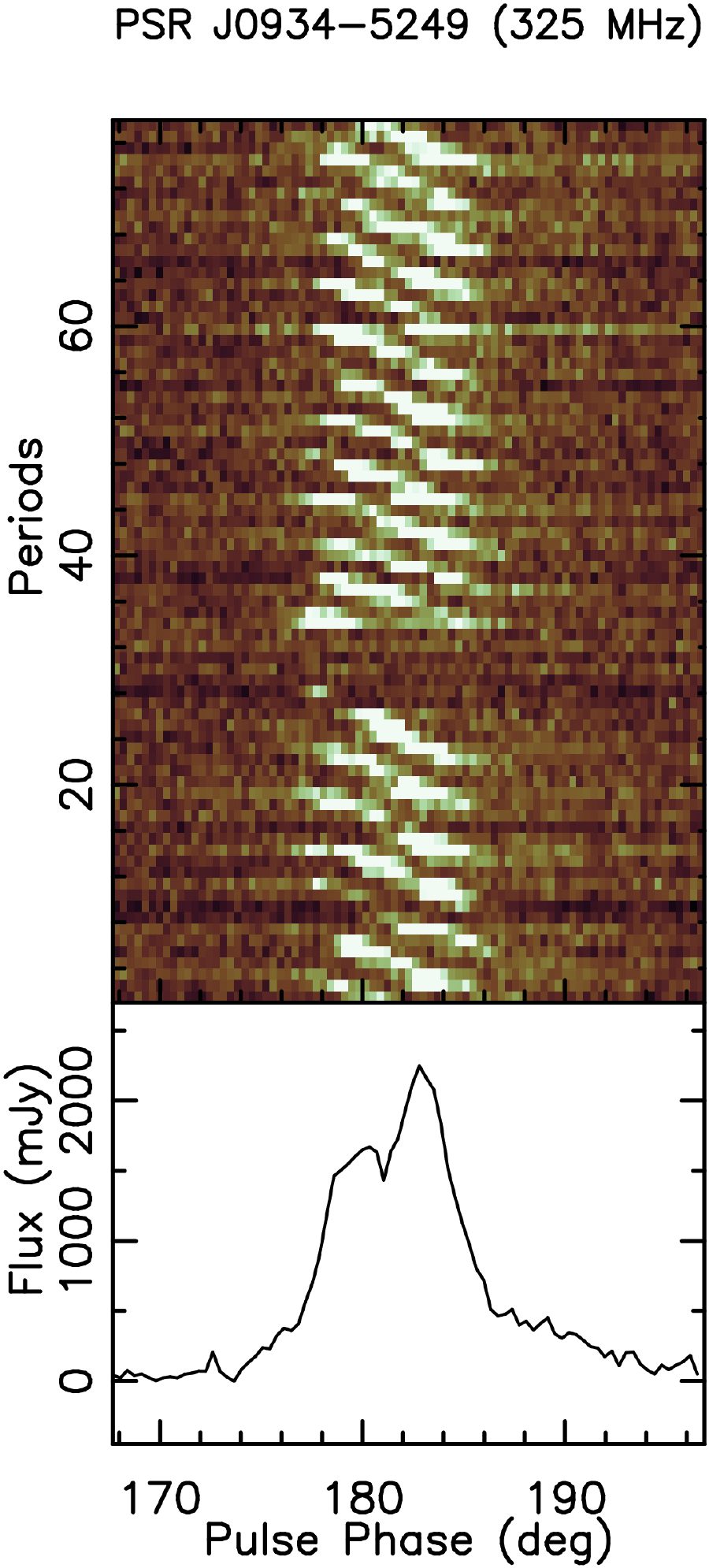}
    \caption{Single pulse sequence in PSR J0934$-$5249 zoomed to show a null.}
    \label{A:0934_sp}
  \end{minipage}
  \hspace{0.5cm}
  \begin{minipage}[b]{0.2\textwidth}
    \includegraphics[width=\textwidth]{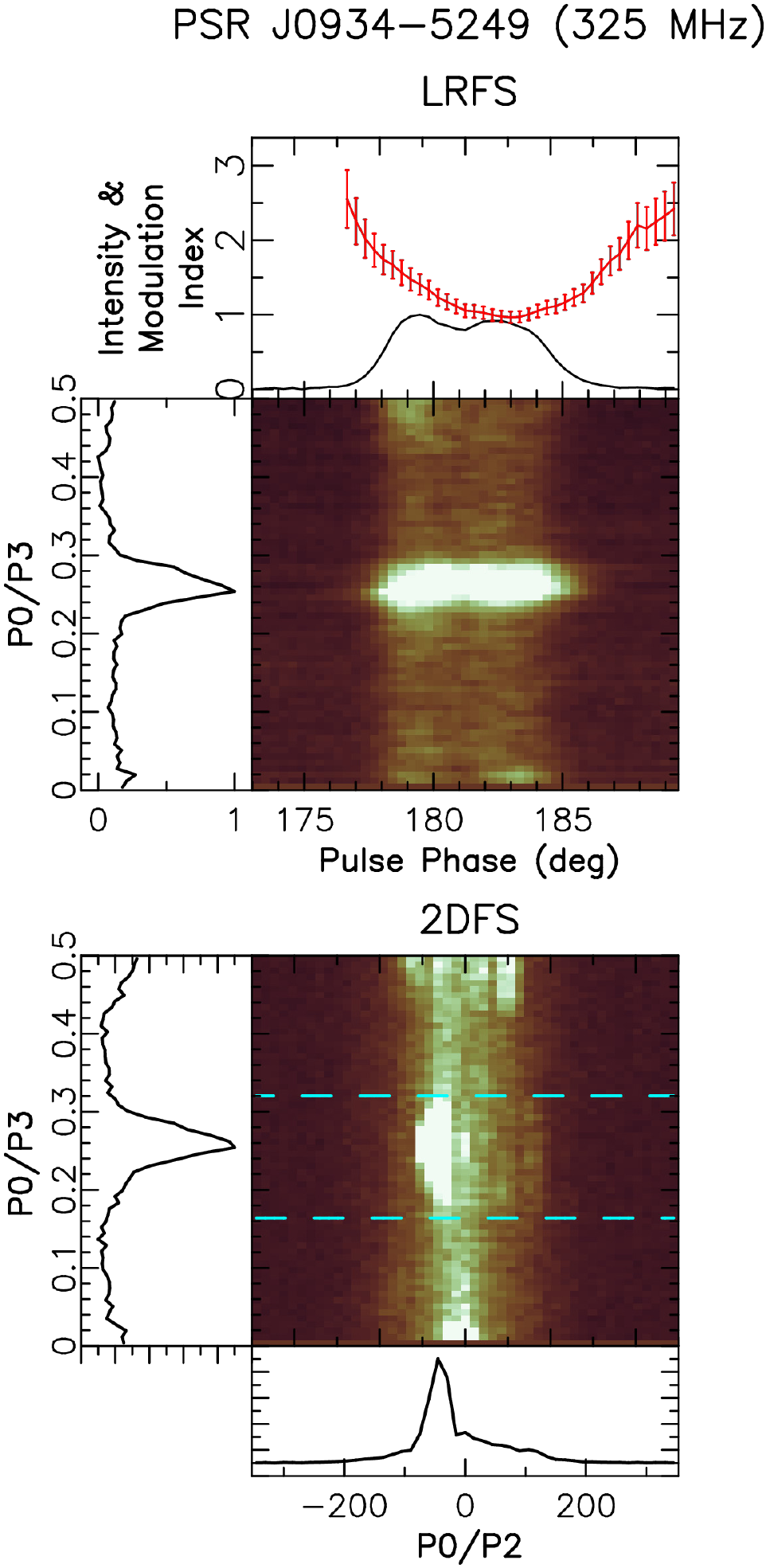}
    \caption{LRFS and 2DFS plots for PSR J0934$-$5249. Bright feature can be seen in the 2DFS plot classifying this pulsar as coherent drifter.}
    \label{A:0934_2dfs}
  \end{minipage}
  \hspace{0.5cm}
  \begin{minipage}[b]{0.2\textwidth}
    \includegraphics[width=\textwidth]{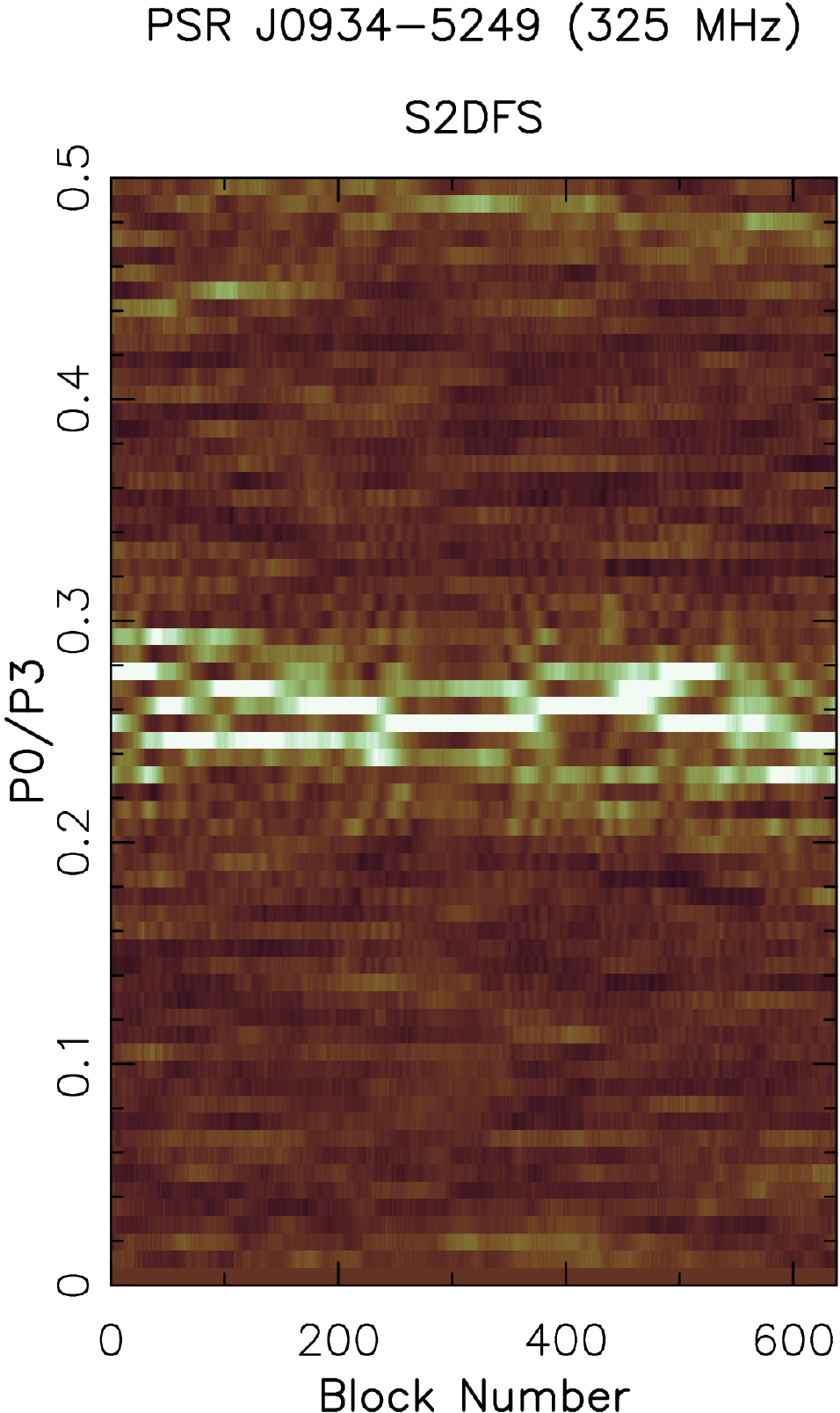}
    \caption{S2DFS of leading component of PSR J0934$-$5249.}
    \label{A:0934_s2dfs}
  \end{minipage}
\end{figure*}

\begin{figure*}[!tbp]
  \centering
  \begin{minipage}[t][][b]{0.45\textwidth}
    \includegraphics[width=\textwidth]{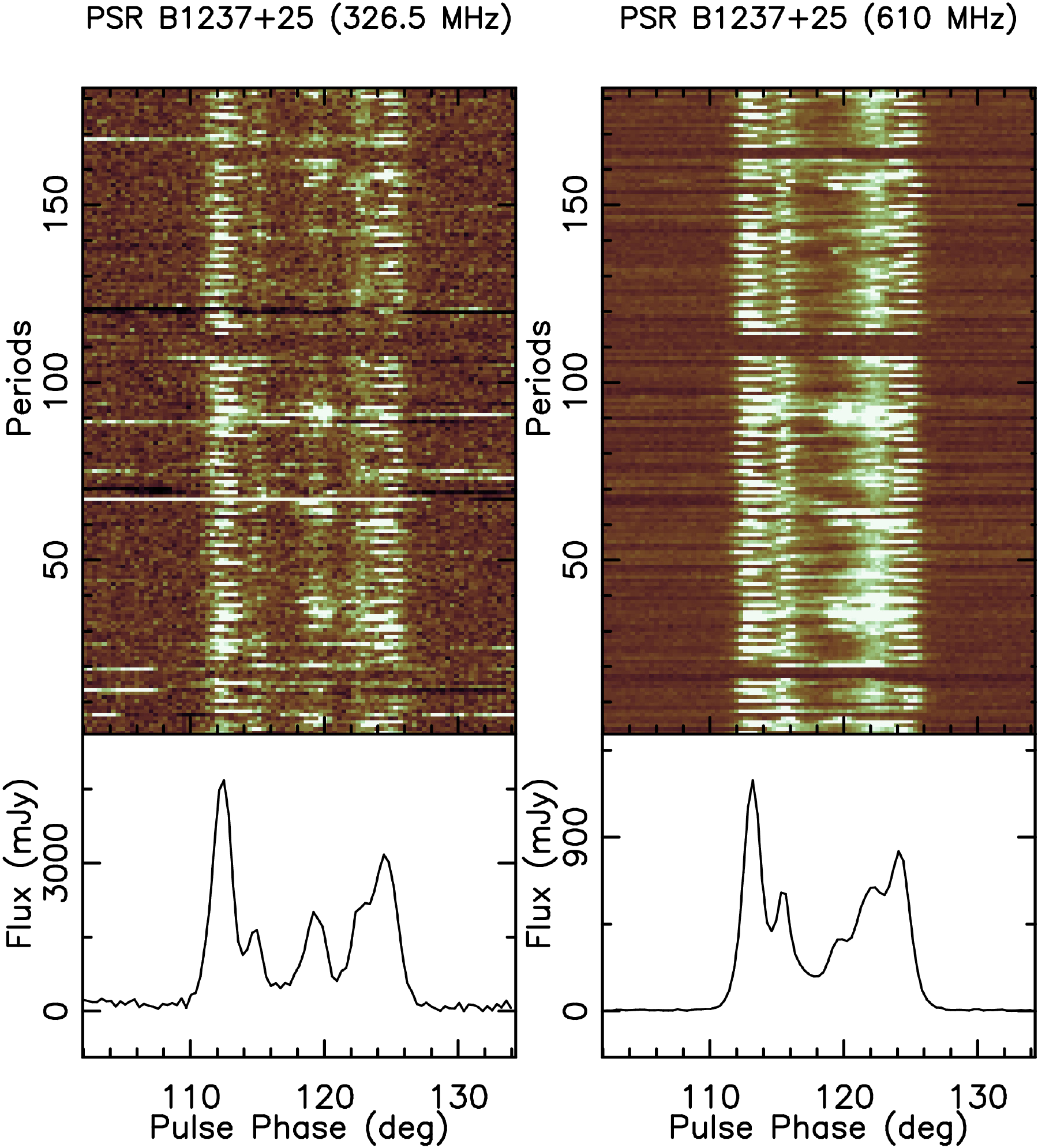}\centering
    \caption{Single pulse sequences in PSR B1237$+$25 zoomed to show a null.}
    \label{A:1237_sp}
 
  \end{minipage}
  \hspace{0.5cm}
  \begin{minipage}[t][][c]{0.45\textwidth}
    \includegraphics[width=\textwidth]{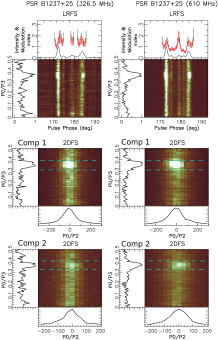}\centering
    \caption{LRFS and 2DFS plots for PSR B1237$+$25. A clear drift feature in both the     trailing and leading components is seen. There is weak feature representing slow drift mode which is difficult to distinguish in this plot. This is more easily seen in Figure \ref{A:1237_2dfs_comp2}}
    \label{A:1237_2dfs}
   
  \end{minipage}
   
\end{figure*}

\clearpage

\begin{figure*}[!tbp]
  \centering
  \begin{minipage}[t][][b]{0.45\textwidth}
    \includegraphics[width=\textwidth]{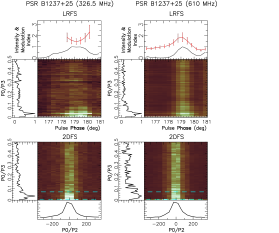}\centering
     \caption{LRFS and 2DFS plots for the second component of PSR B1237$+$25. A low frequency drift feature can be seen in the LRFS at 326.5 MHz. A weak feature is also seen in the LRFS at  610 MHz.}
     \label{A:1237_2dfs_comp2}
  
  \end{minipage}
  \hspace{0.5cm}
  \begin{minipage}[t][][c]{0.45\textwidth}
    \includegraphics[width=\textwidth]{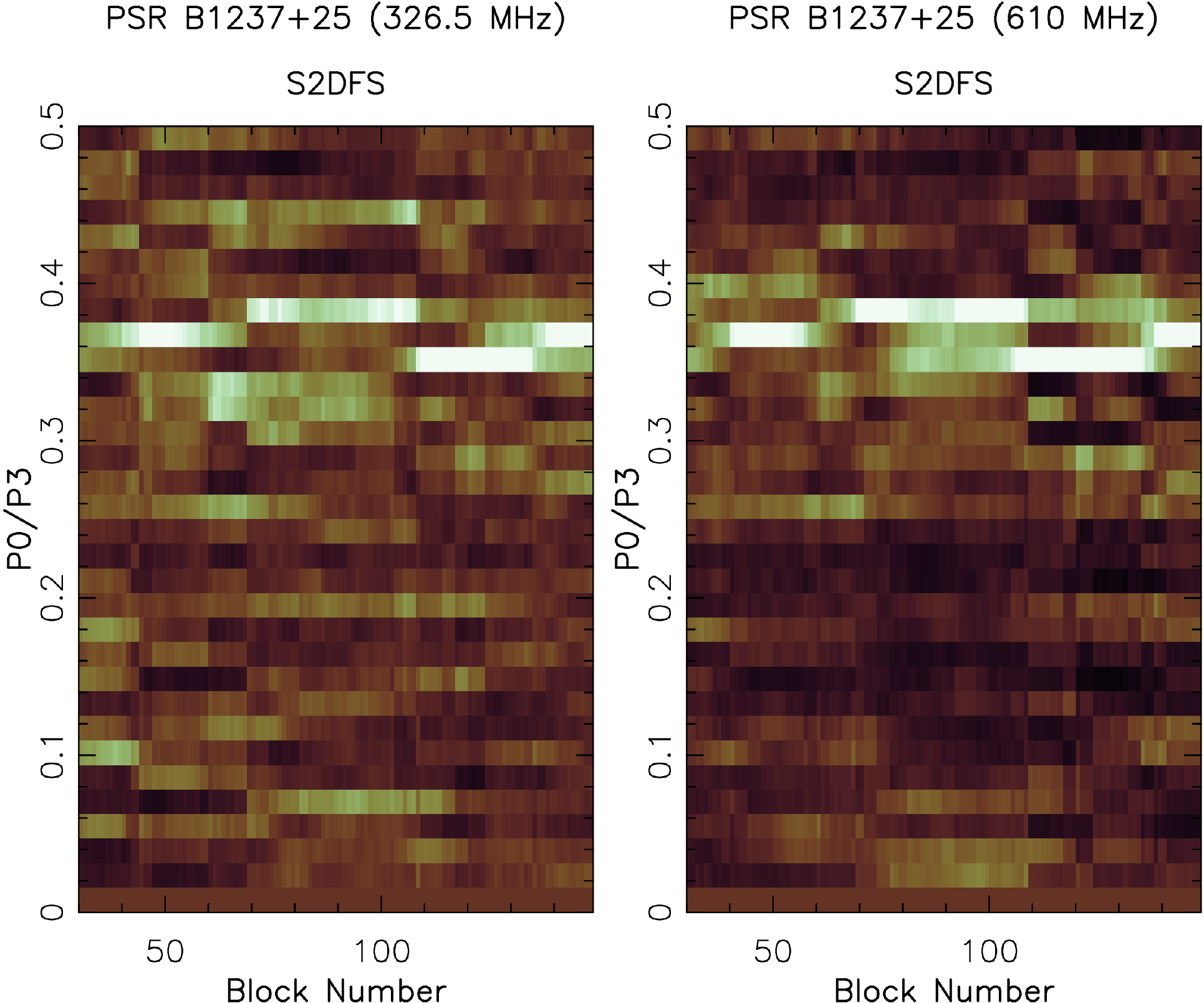}\centering
    \caption{S2DFS of leading component of PSR B1237$+$25.}
    \label{A:1237_s2dfs}
         
  \end{minipage}
   
\end{figure*}

\begin{figure*}[!tbp]
  \centering
  \begin{minipage}[t][][b]{0.45\textwidth}
    \includegraphics[width=\textwidth]{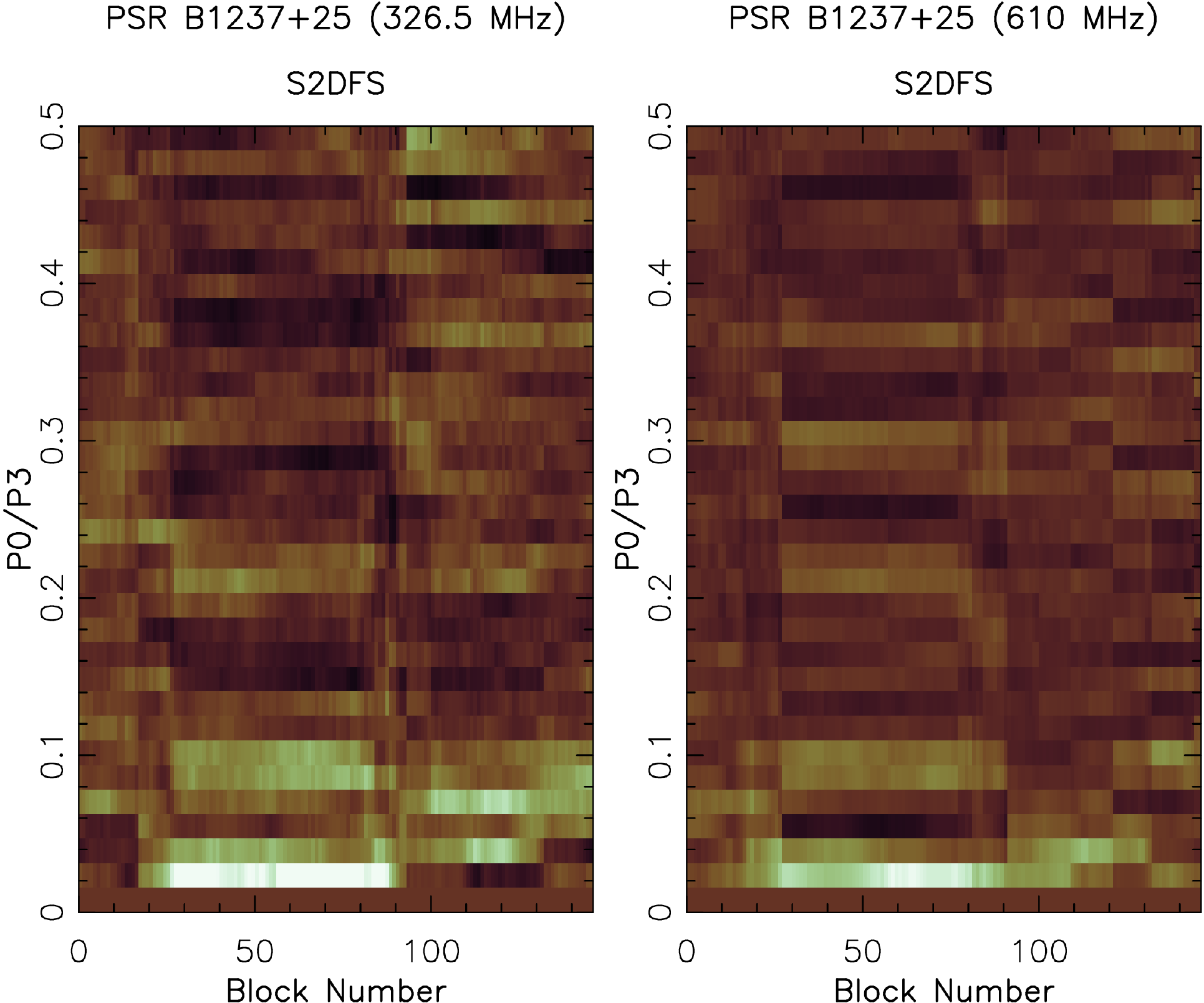}\centering
     \caption{S2DFS of central component of PSR B1237$+$25.}
     \label{A:1237_s2dfs_1}
  
  \end{minipage}
  \hspace{0.5cm}
  \begin{minipage}[t][][c]{0.45\textwidth}
    \includegraphics[width=\textwidth]{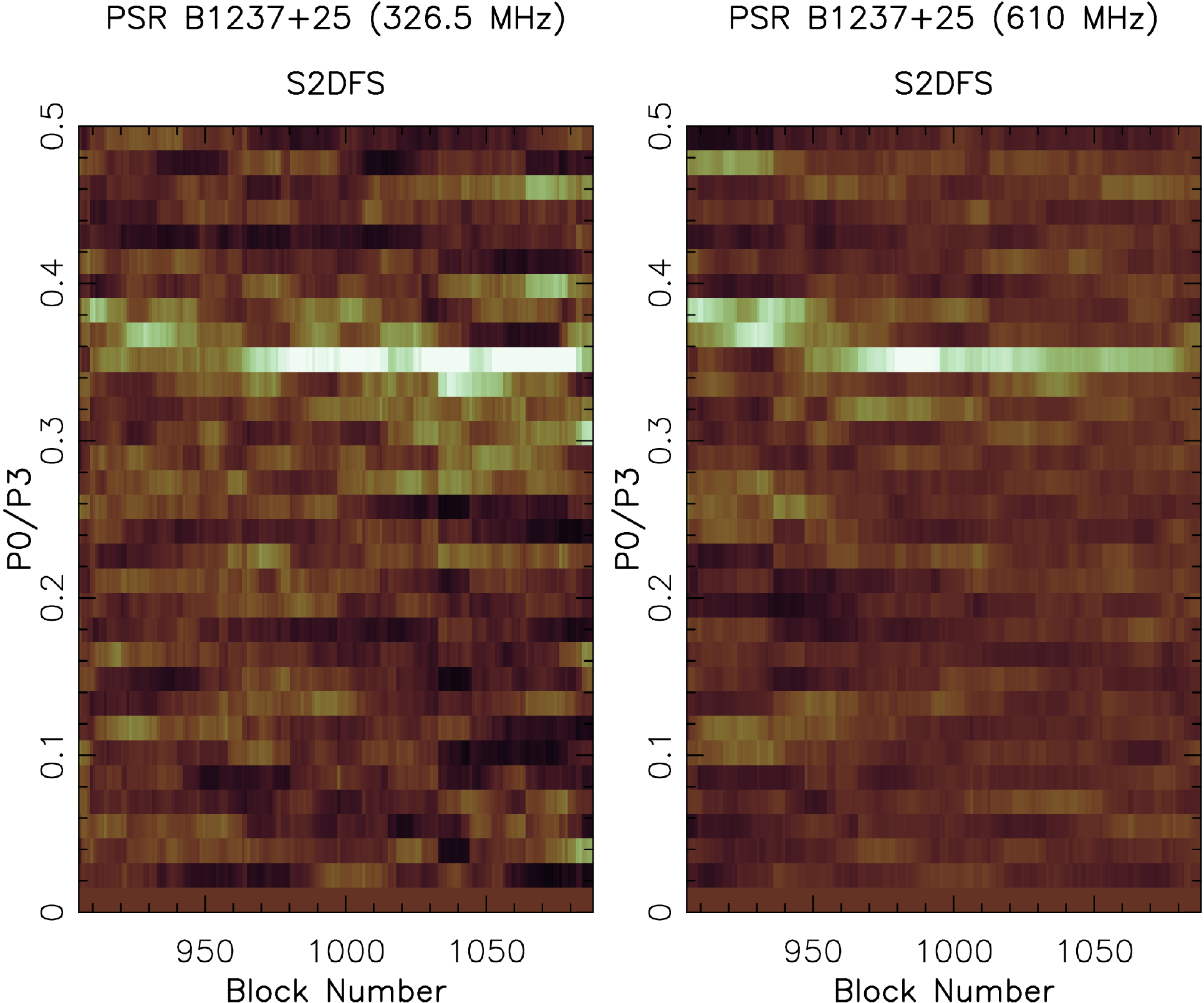}\centering
    \caption{S2DFS of trailing component of PSR B1237$+$25.}
    \label{A:1237_s2dfs_2}
         
  \end{minipage}
   
\end{figure*}

\begin{figure*}
\includegraphics[width=0.45\textwidth]{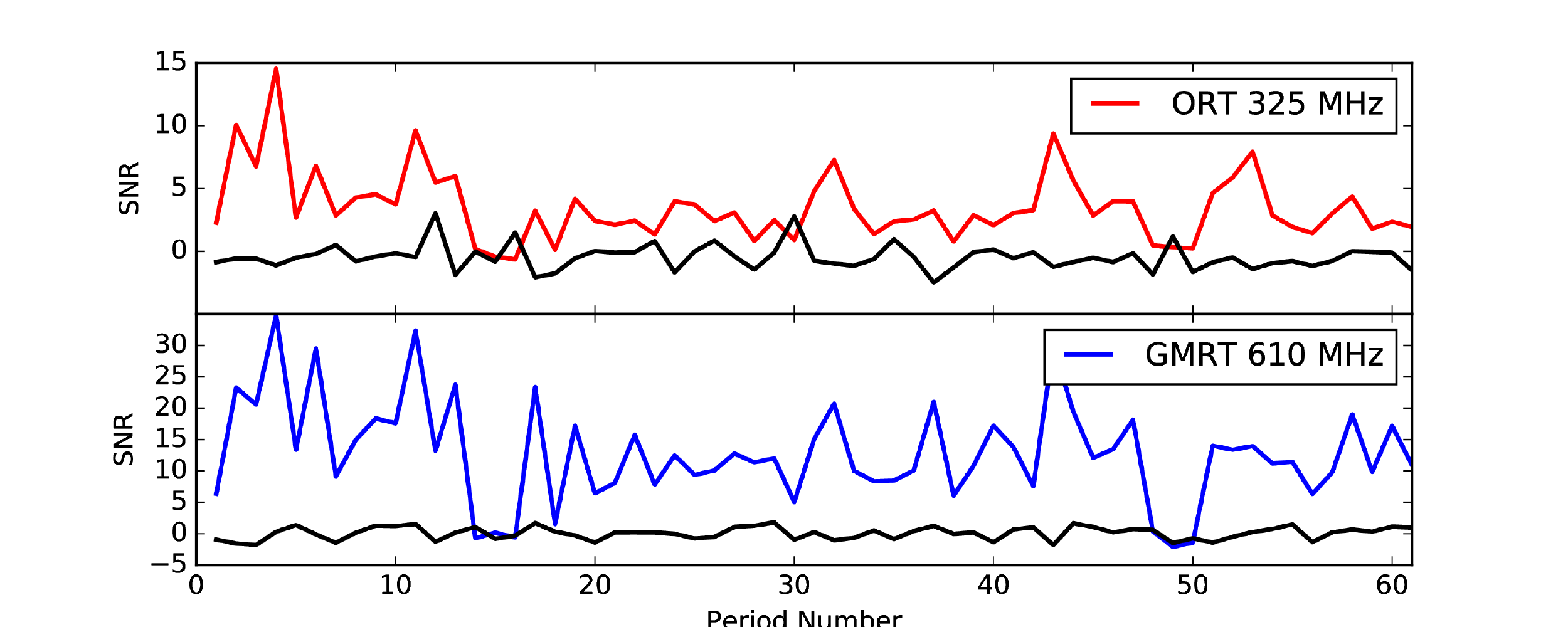}\centering
\caption{On-pulse energy sequence of PSR B1237$+$25.}
\label{A:1237_ep}
\end{figure*}

\clearpage
 
\begin{figure*}[!tbp]
  \centering
  \begin{minipage}[t][][b]{0.45\textwidth}
    \includegraphics[width=\textwidth]{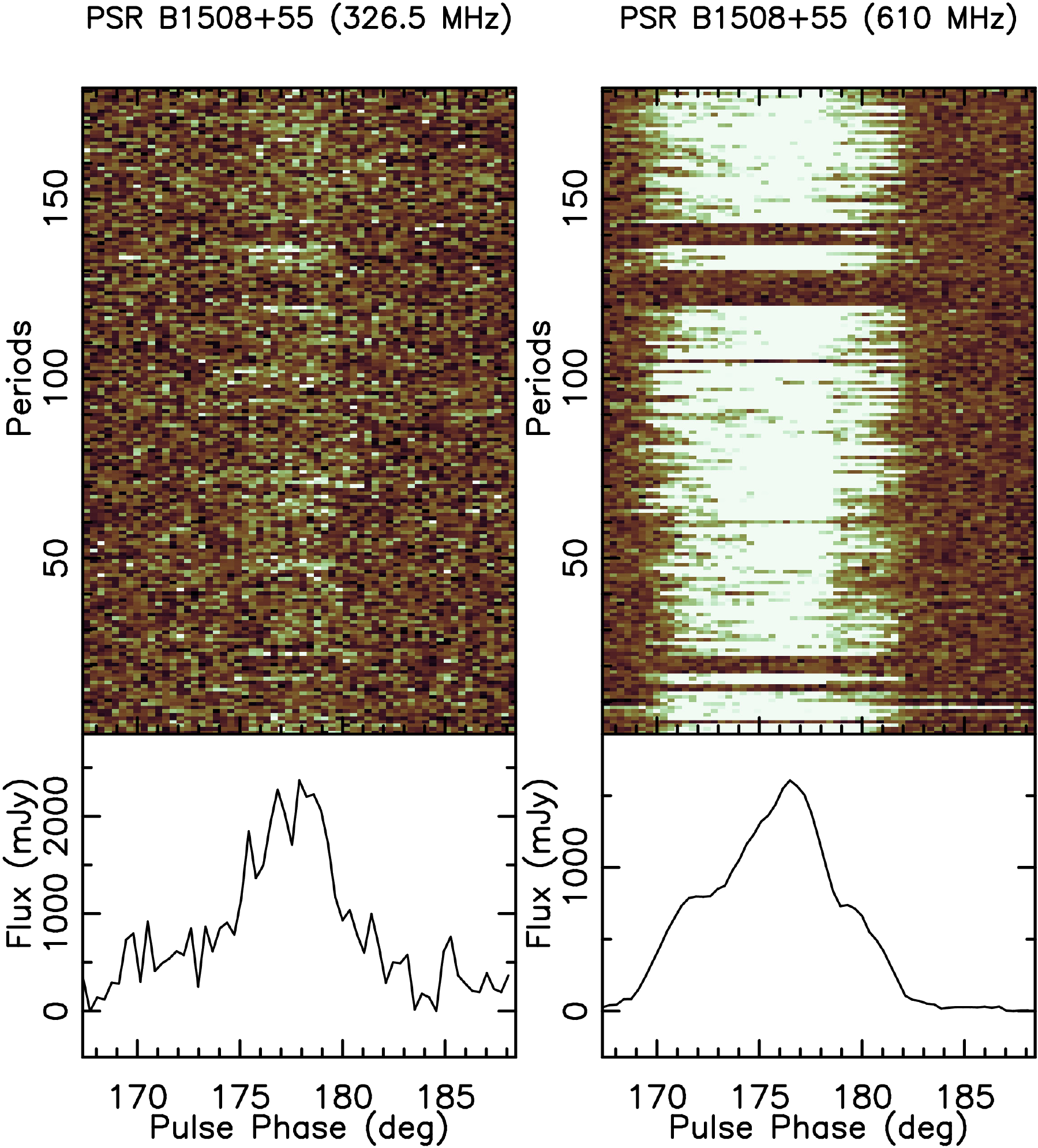}\centering
   \caption{Single pulse sequence for PSR B1508$-$55.}
   \label{A:1508_sp}
     
  \end{minipage}
  \hspace{0.5cm}
  \begin{minipage}[t][][c]{0.45\textwidth}
    \includegraphics[width=\textwidth]{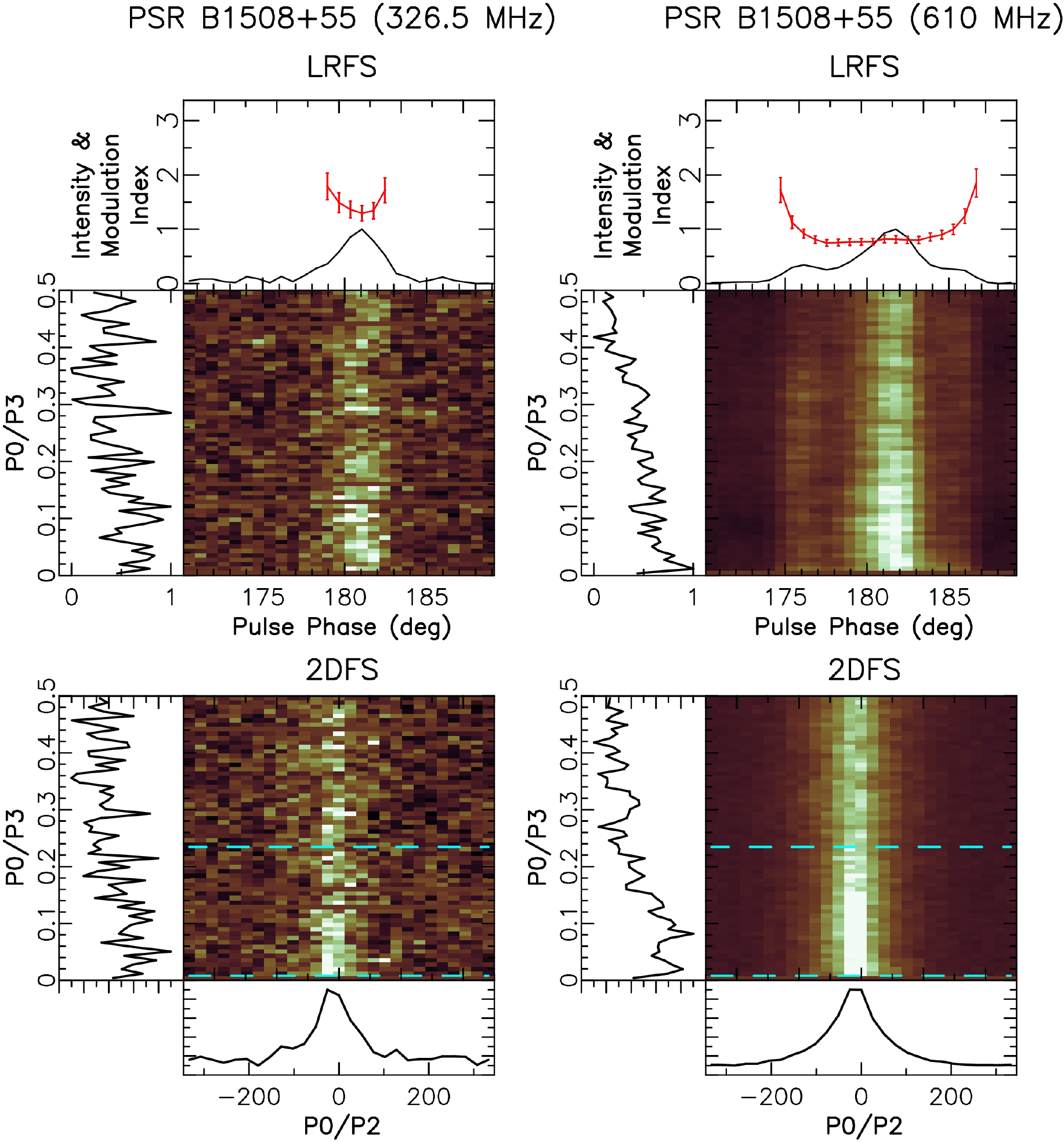}\centering
   \caption{LRFS and 2DFS plots for PSR B1508$+$55. A broad low frequency feature in both the frequencies is seen.}
   \label{A:1508_2dfs}  
   
  \end{minipage}
   
\end{figure*} 

\begin{figure*}[!tbp]
  \centering
  \begin{minipage}[c][][b]{0.45\textwidth}
    \includegraphics[width=\textwidth]{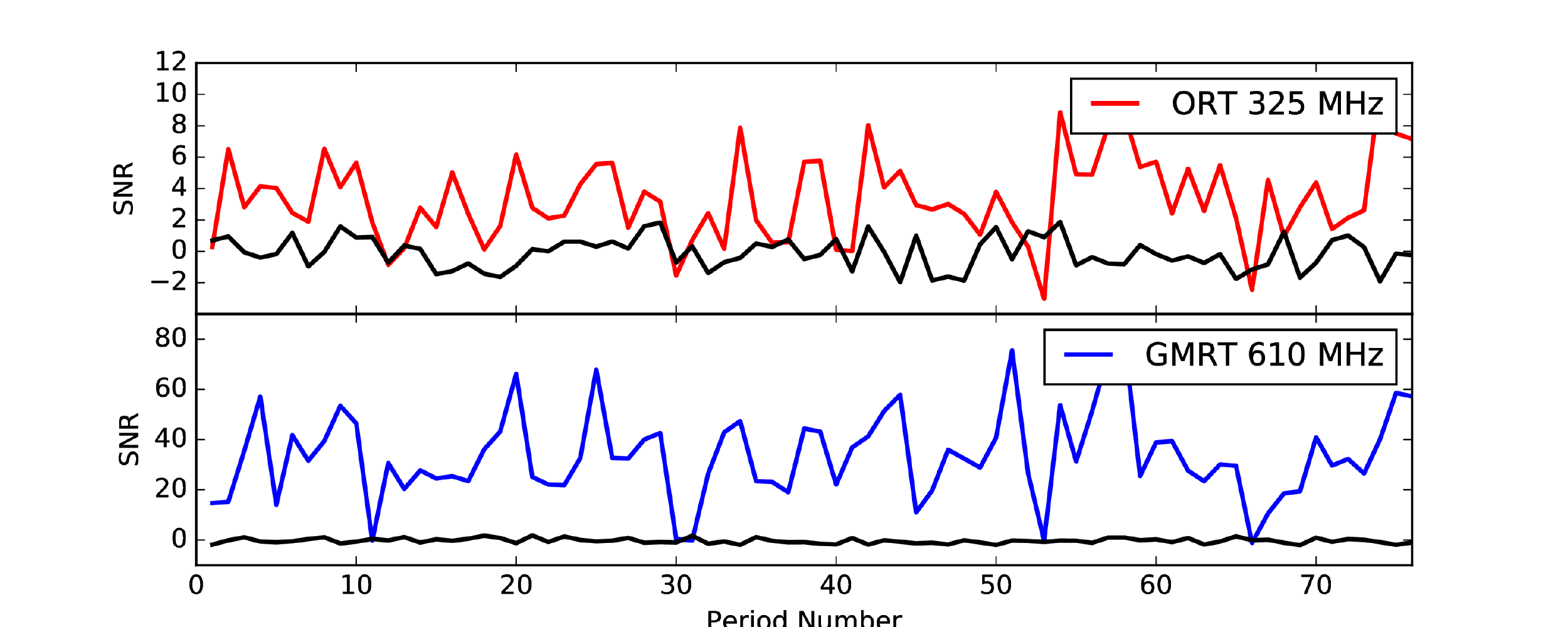}\centering
    \caption{On-pulse energy sequence of PSR B1508$-$06.}
    \label{A:1508_ep}

  \end{minipage}
  \hspace{0.5cm}
  \begin{minipage}[c][][b]{0.45\textwidth}
   \includegraphics[width=\textwidth]{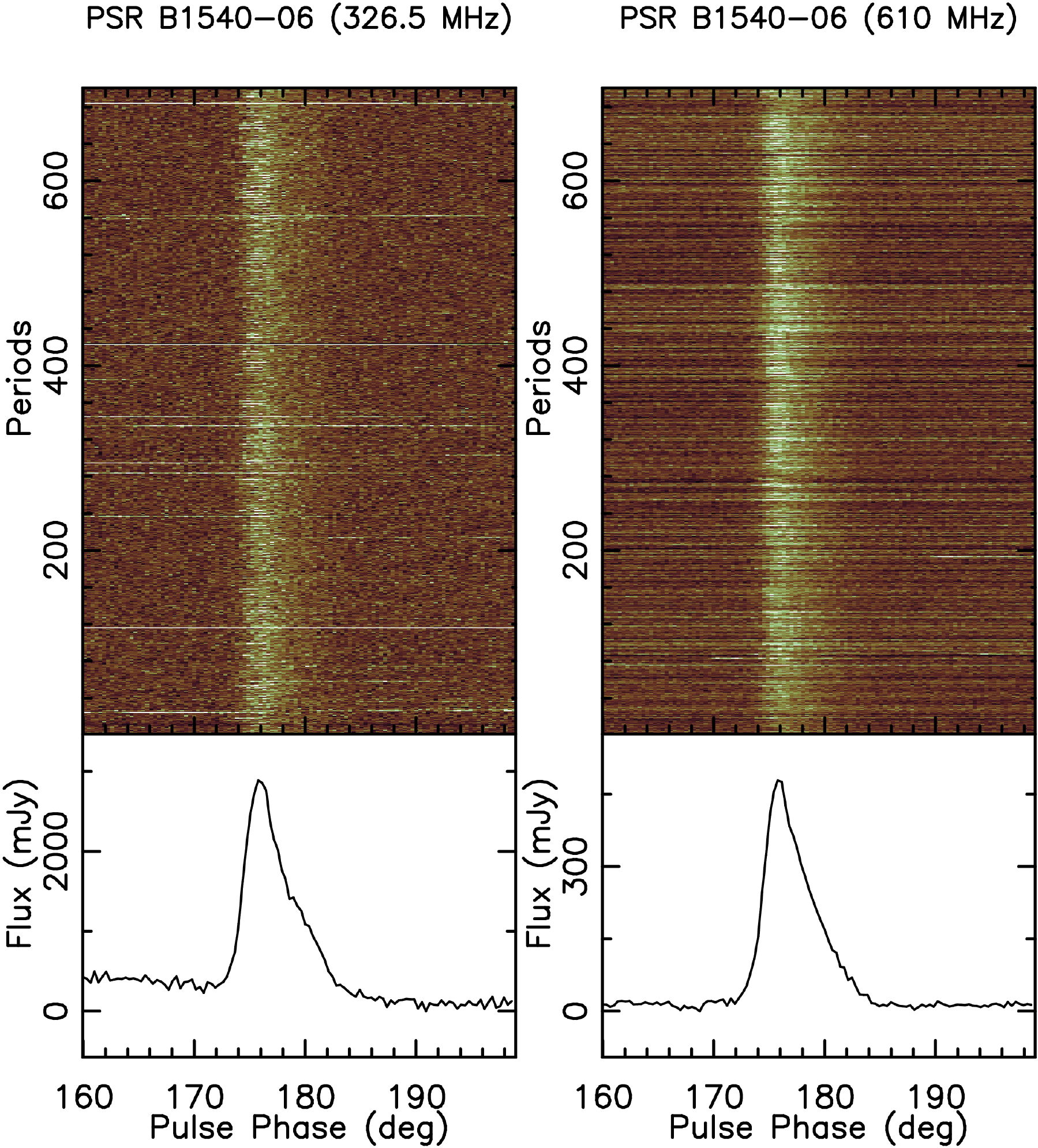}\centering
\caption{Single pulse sequence for PSR B1540$-$06.}
   \label{A:1540_sp}        
  \end{minipage}
   
\end{figure*}  

\begin{figure*}[!tbp]
  \centering
  \begin{minipage}[c][][b]{0.45\textwidth}
    \includegraphics[width=\textwidth]{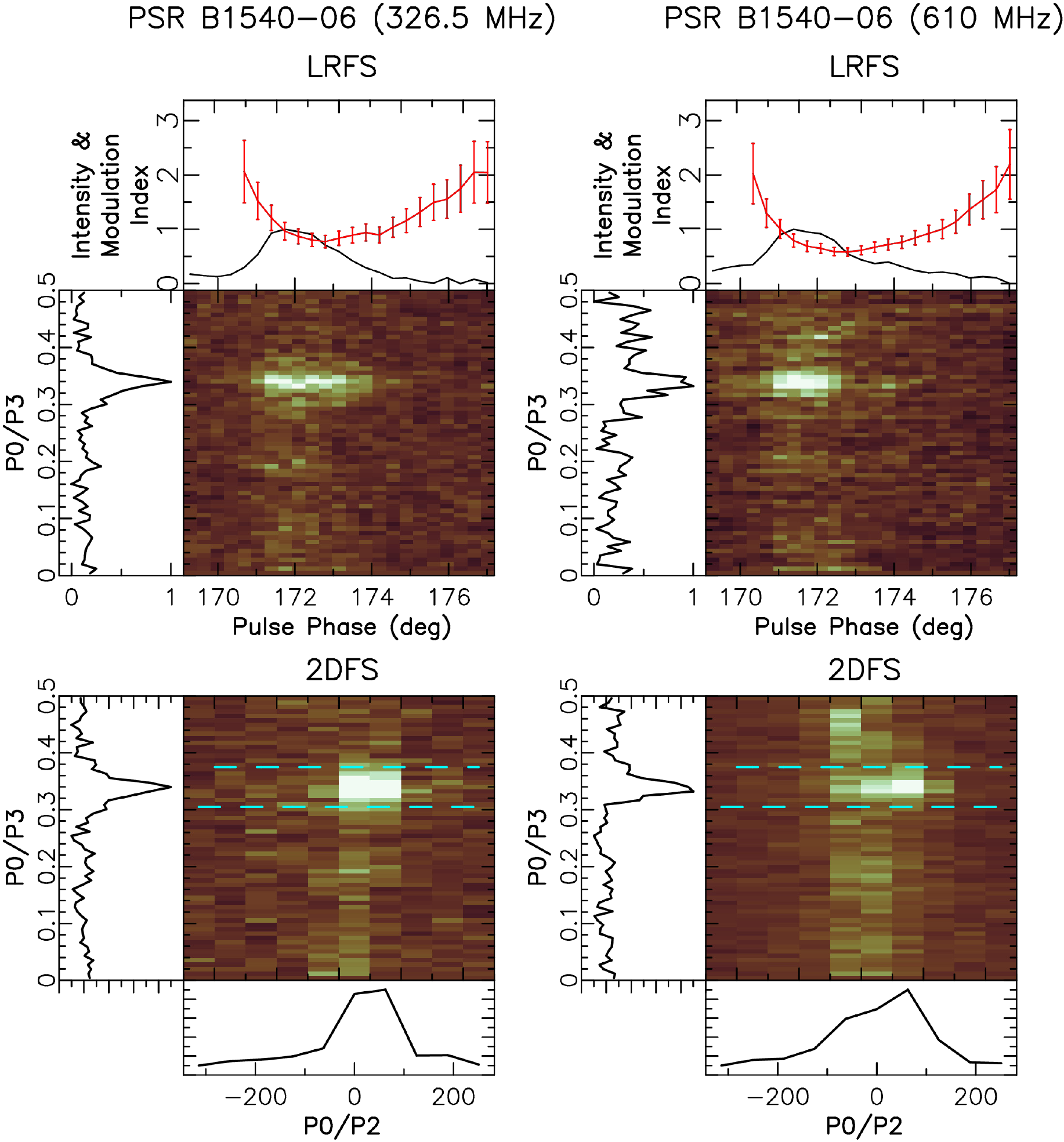}\centering
    \caption{LRFS and 2DFS plots for B1540$-$06. A single bright feature at both the frequencies is seen.}
    \label{A:1540_2dfs}

  \end{minipage}
  \hspace{0.5cm}
  \begin{minipage}[c][][b]{0.45\textwidth}
   \includegraphics[width=\textwidth]{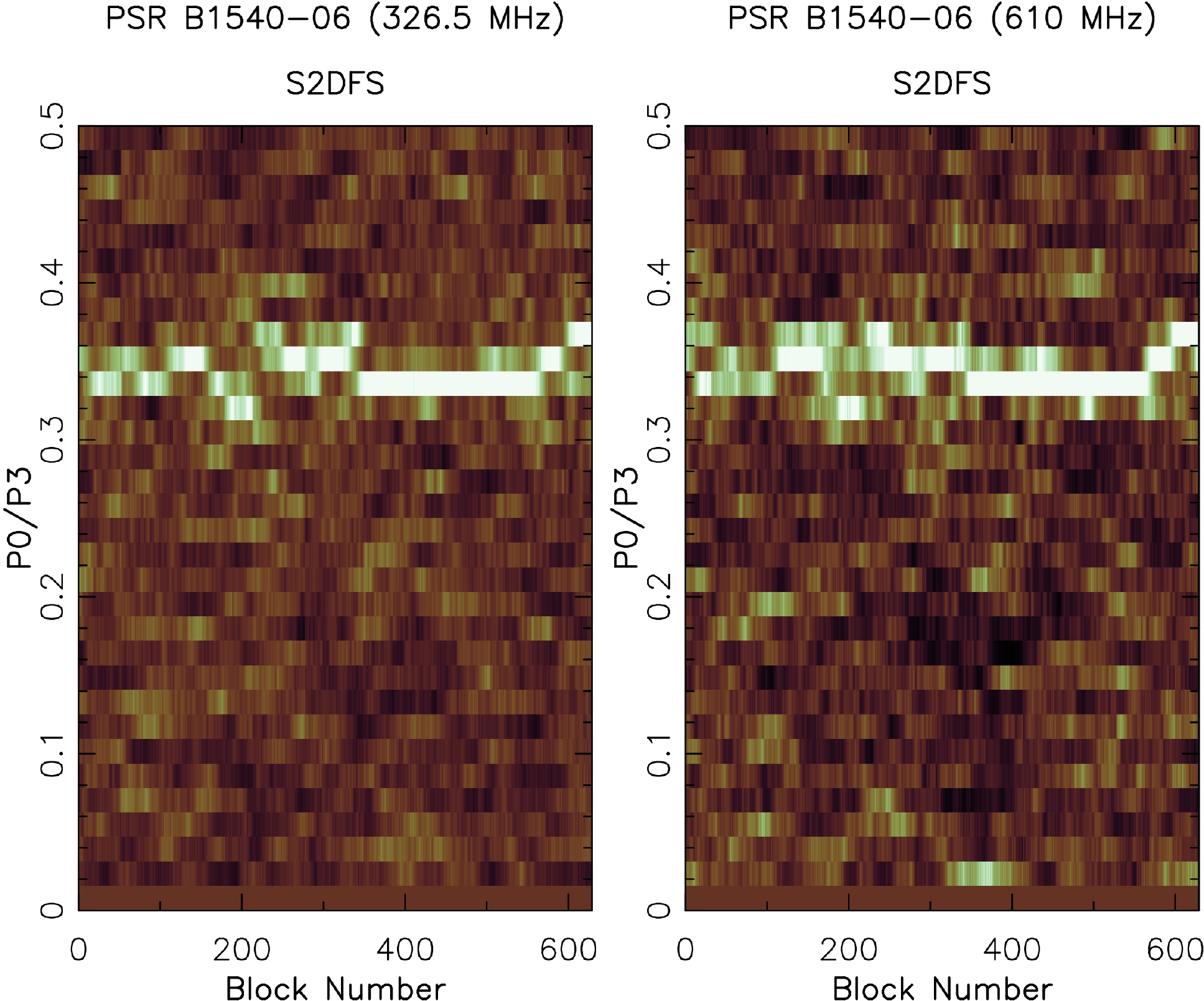}\centering
\caption{S2DFS of PSR B1540$-$06.}
   \label{A:1540_s2dfs}        
  \end{minipage}
   
\end{figure*}


\begin{figure*}[!tbp]
  \centering
  \begin{minipage}[b]{0.2\textwidth}
    \includegraphics[width=\textwidth]{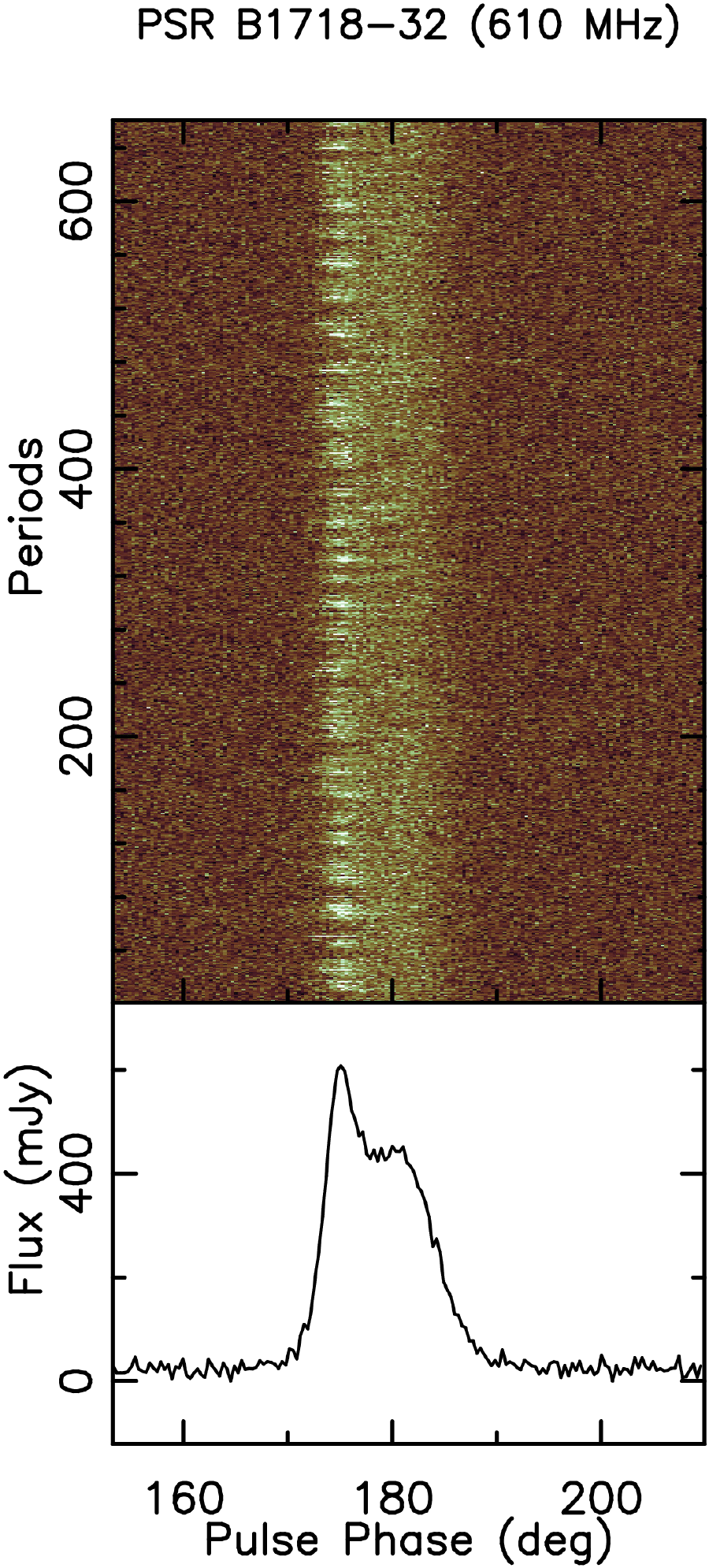}
    \caption{Single pulse sequence for PSR B1718$-$32. The single pulse SNR is low at 326.5 MHz and 1308 MHz observations.}
    \label{A:1718_sp}
  \end{minipage}
  \hspace{0.5cm}
  \begin{minipage}[b]{0.2\textwidth}
    \includegraphics[width=\textwidth]{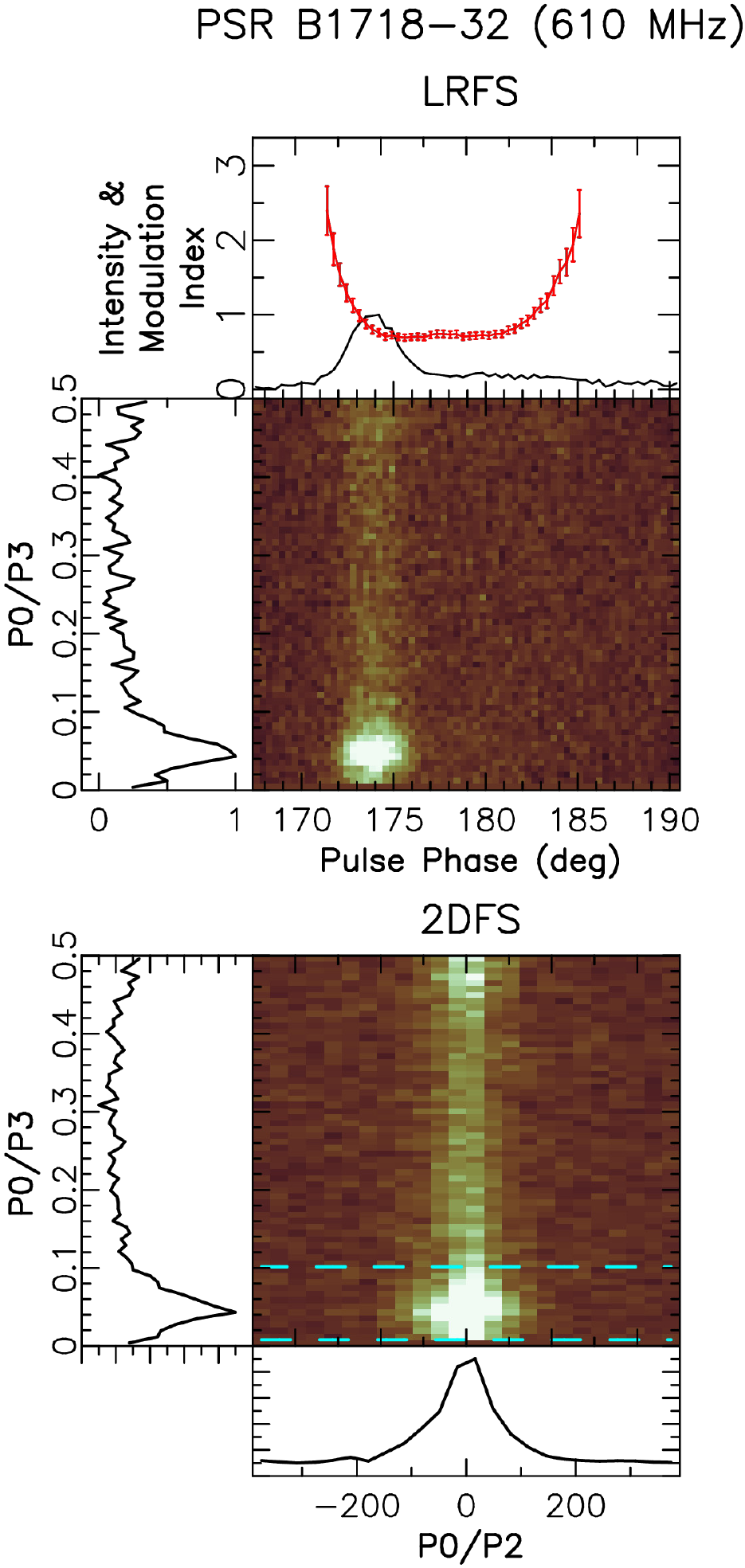}
    \caption{LRFS and 2DFS plots for PSR B1718$-$32. Strong drift feature is seen only at 610 MHz. Non detection at other frequencies  is probably due to the low SNR.}
    \label{A:1718_2dfs}
  \end{minipage}
  \hspace{0.5cm}
  \begin{minipage}[b]{0.2\textwidth}
    \includegraphics[width=\textwidth]{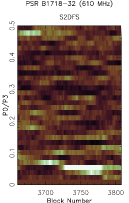}
    \caption{S2DFS of PSR B1718$-$32.}
    \label{A:1718_s2dfs}
  \end{minipage}
\end{figure*}

\begin{figure*}[!tbp]
  \centering
  \begin{minipage}[b]{0.2\textwidth}
    \includegraphics[width=\textwidth]{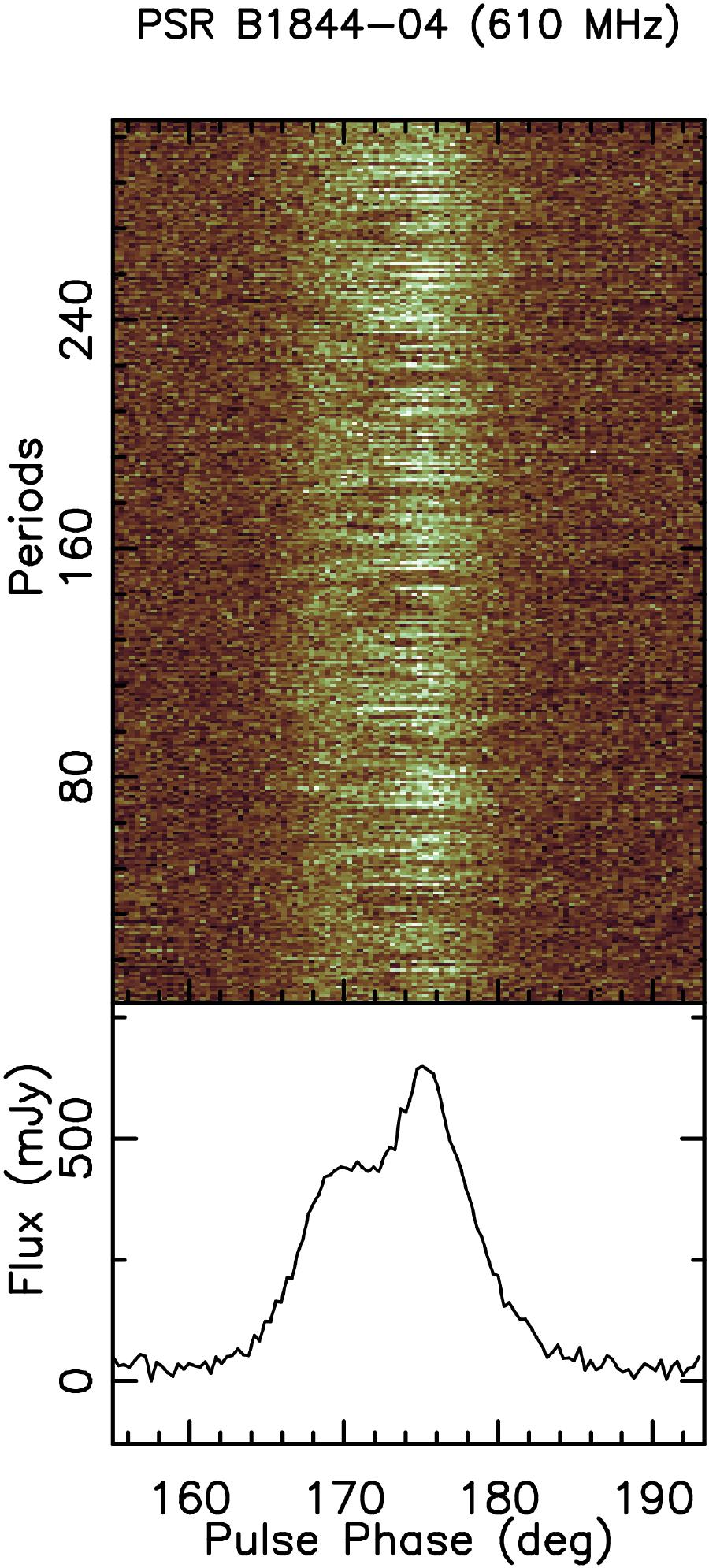}
    \caption{Single pulse sequences for PSR B1844$-$04. The single pulse SNR is low for 326.5 MHz observation.}
    \label{A:1844_sp}
  \end{minipage}
  \hspace{0.5cm}
  \begin{minipage}[b]{0.2\textwidth}
    \includegraphics[width=\textwidth]{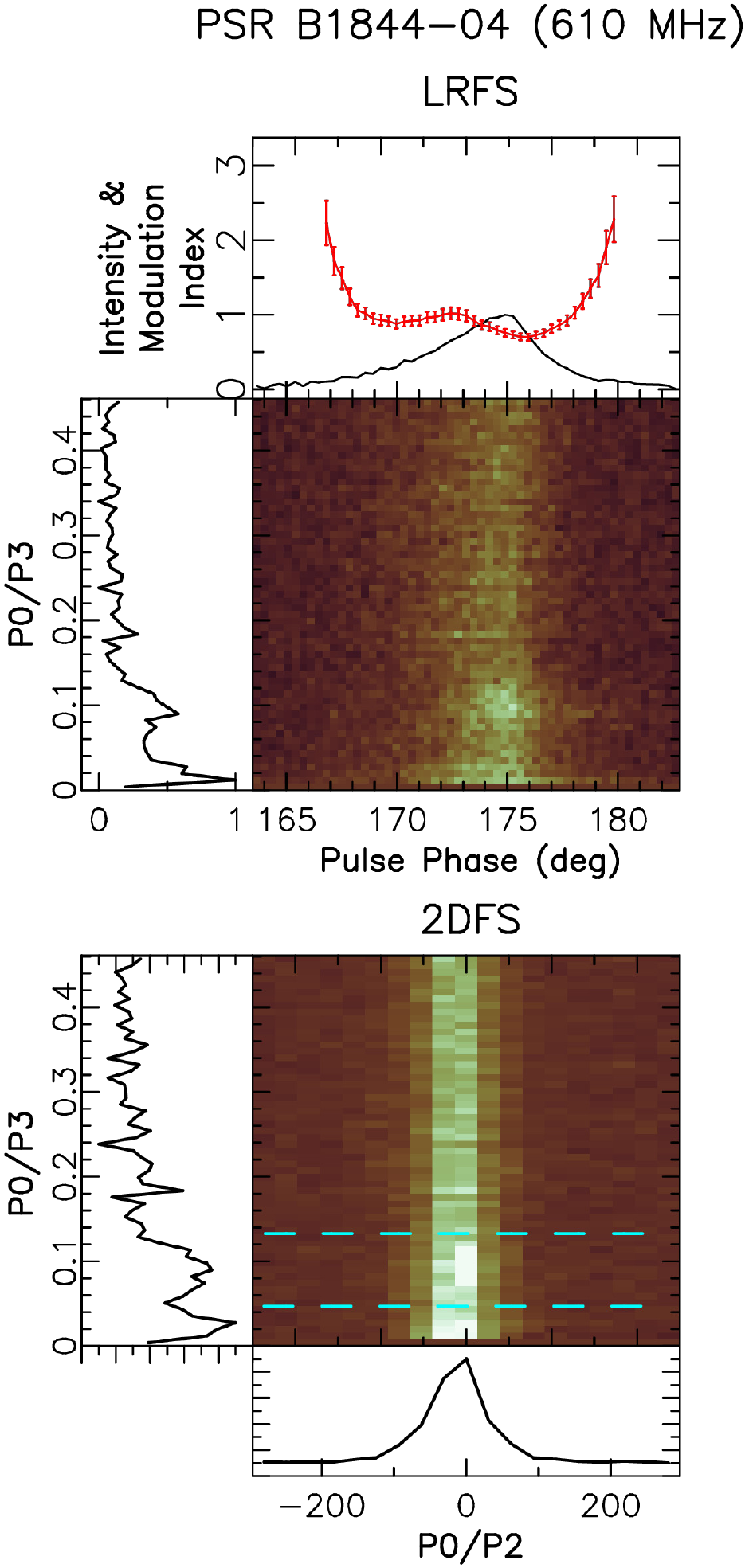}
    \caption{LRFS and 2DFS plots for PSR B1844$-$04. The plots show a broad feature at 610 MHz.}
    \label{A:1844_2dfs}
  \end{minipage}
  \hspace{0.5cm}
  \begin{minipage}[b]{0.2\textwidth}
    \includegraphics[width=\textwidth]{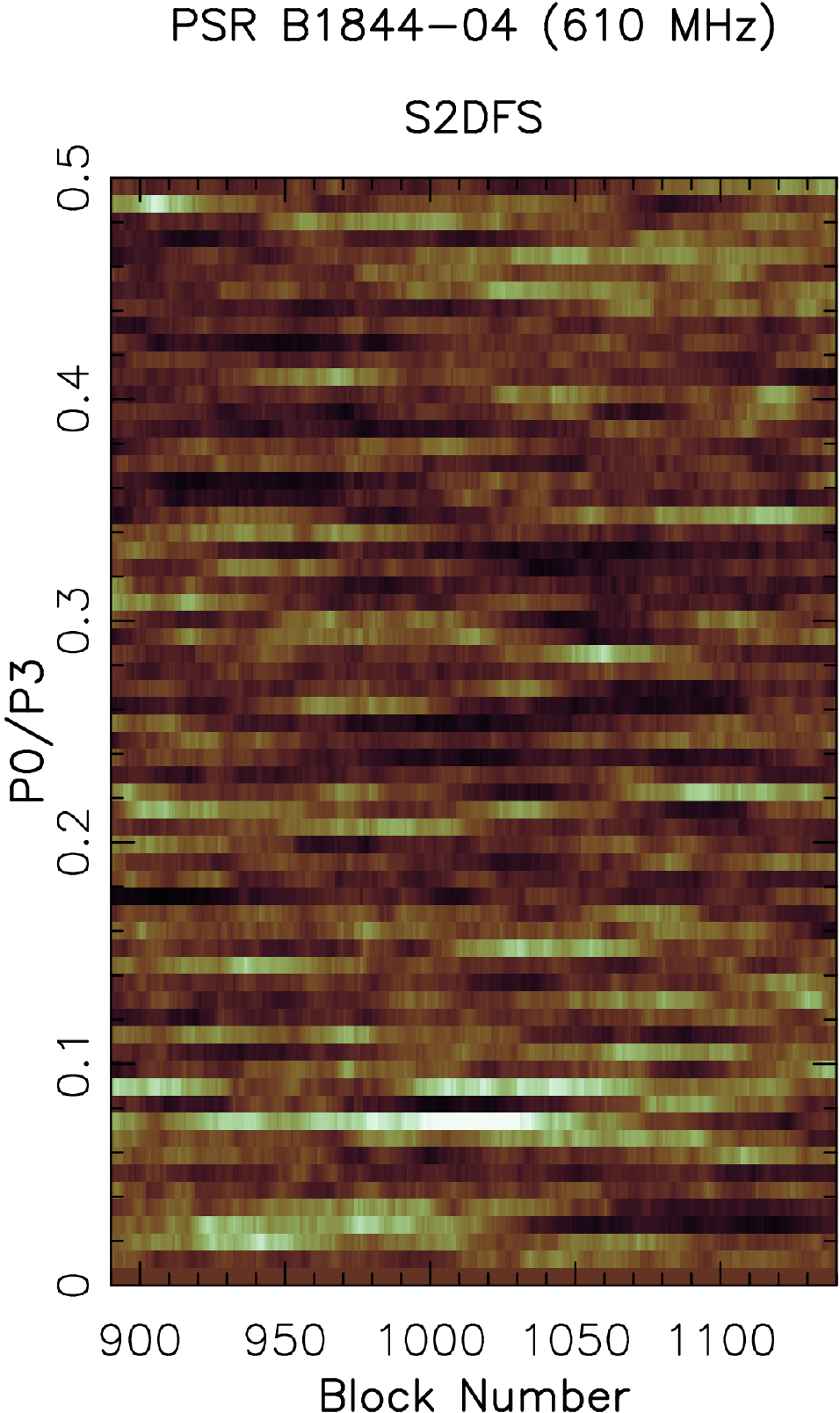}
    \caption{S2DFS of PSR B1844$-$04.}
    \label{A:1844_s2dfs}
  \end{minipage}
\end{figure*} 

\begin{figure*}[!tbp]
  \centering
  \begin{minipage}[c][][b]{0.45\textwidth}
    \includegraphics[width=\textwidth]{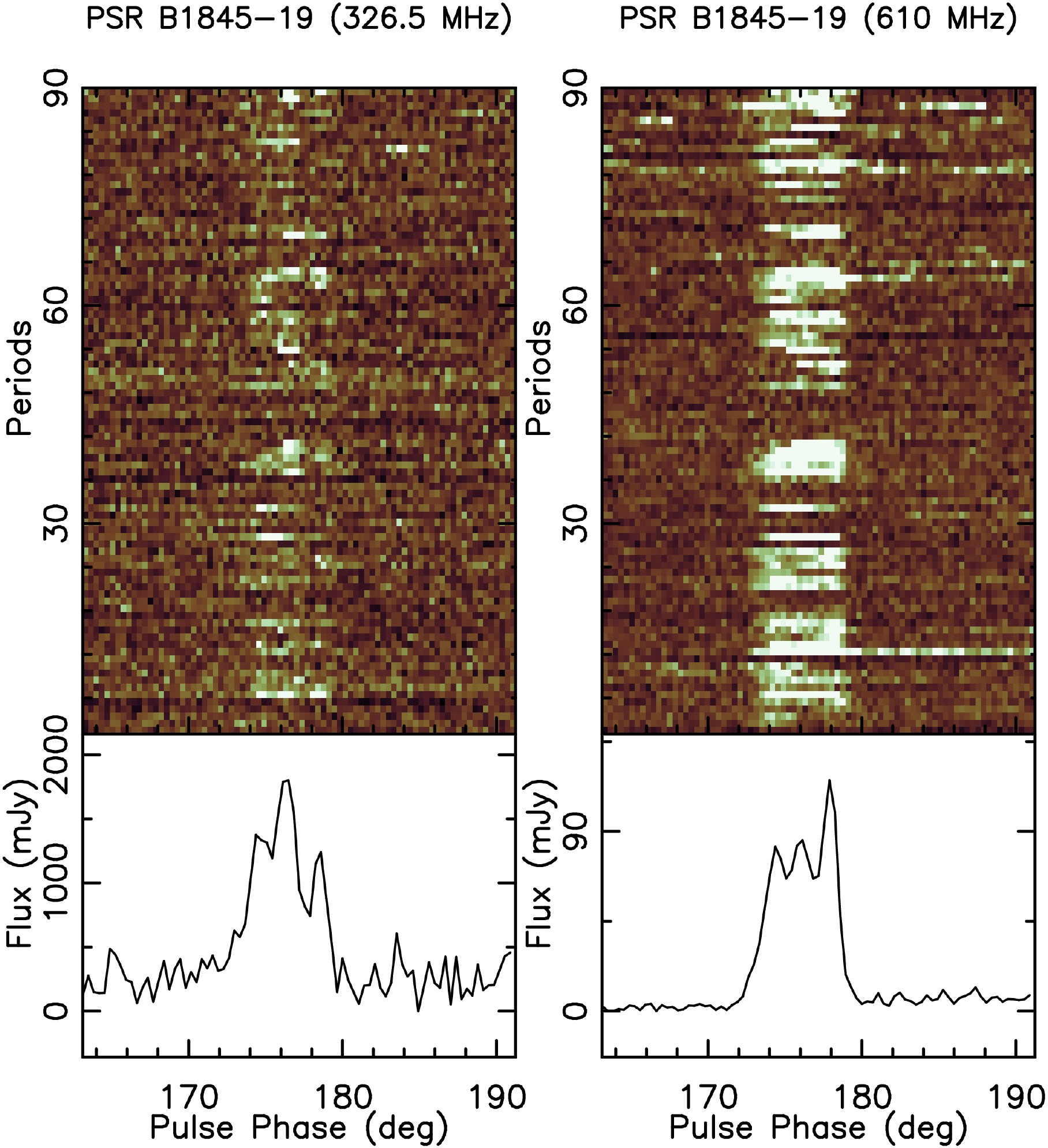}\centering
    \caption{Single pulse sequences for PSR B1845$-$19.}
    \label{A:1845_sp}

  \end{minipage}
  \hspace{0.5cm}
  \begin{minipage}[c][][b]{0.45\textwidth}
   \includegraphics[width=\textwidth]{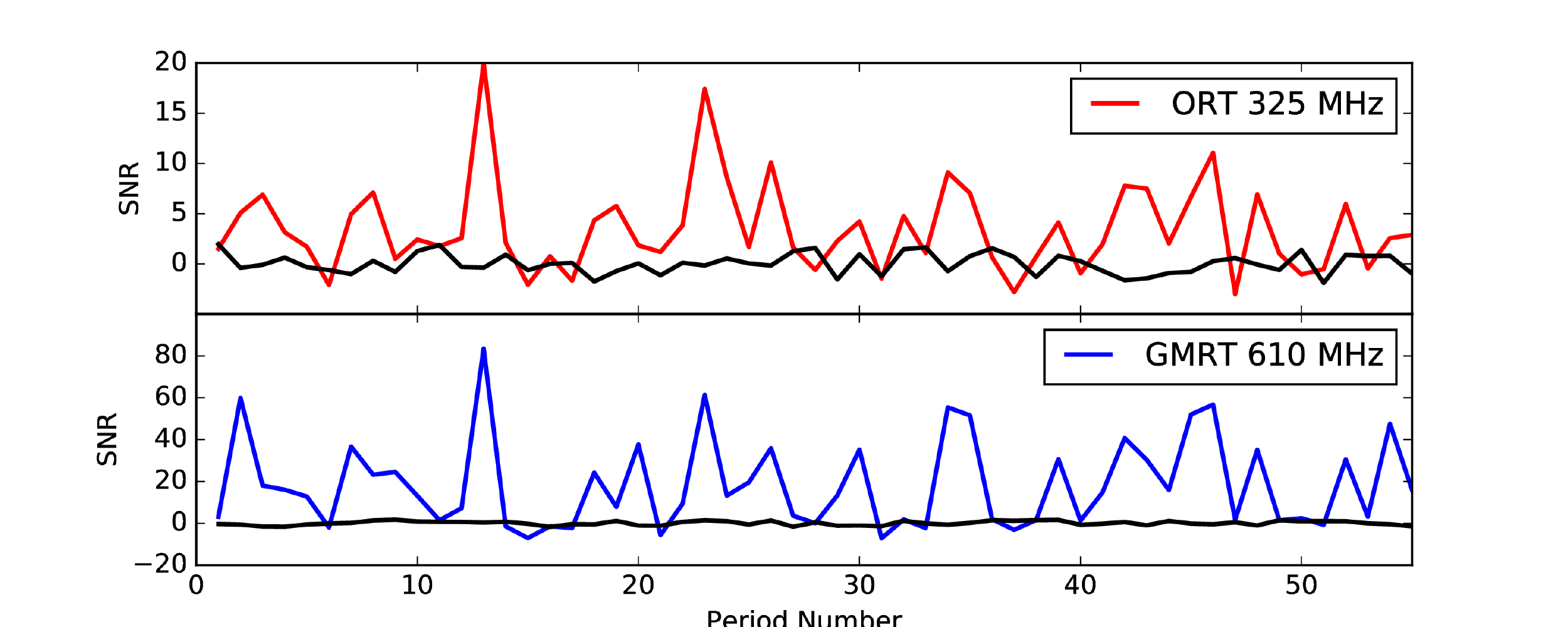}\centering
\caption{On-pulse energy sequence of PSR B1845$+$55.}
   \label{A:1845_ep}        
  \end{minipage}
   
\end{figure*}

 
\begin{figure*}[!tbp]
  \centering
  \begin{minipage}[c][][b]{0.45\textwidth}
    \includegraphics[width=\textwidth]{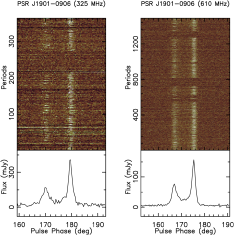}\centering
    \caption{Single pulse sequences of independent observations at 325 MHz and 610 MHz using the GMRT for PSR J1901$-$0906.}
    \label{A:1901_sp}

  \end{minipage}
  \hspace{0.5cm}
  \begin{minipage}[c][][b]{0.45\textwidth}
   \includegraphics[width=\textwidth]{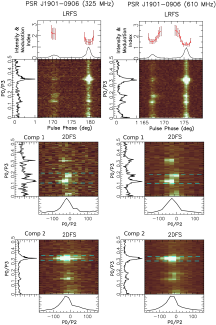}\centering
\caption{LRFS and 2DFS plots for PSR J1901$-$0906. Three three clear drift features are seen in both the components. Fast mode is more prominent in the trailing component.}
   \label{A:1901_2dfs}        
  \end{minipage}
   
\end{figure*}  


\begin{figure*}[!tbp]
  \centering
   
  \begin{minipage}[c][][b]{0.45\textwidth}
   \includegraphics[width=\textwidth]{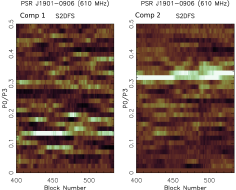}\centering
\caption{S2DFS for the two components of PSR J1901$-$0906 observed at 610 MHz. The plot on left is the S2DFS for the leading component and the plot on the right is  for the  trailing component.}
   \label{A:1901_s2dfs}        
  \end{minipage}
   
\end{figure*}

\begin{figure*}
\includegraphics[width=0.60\textwidth]{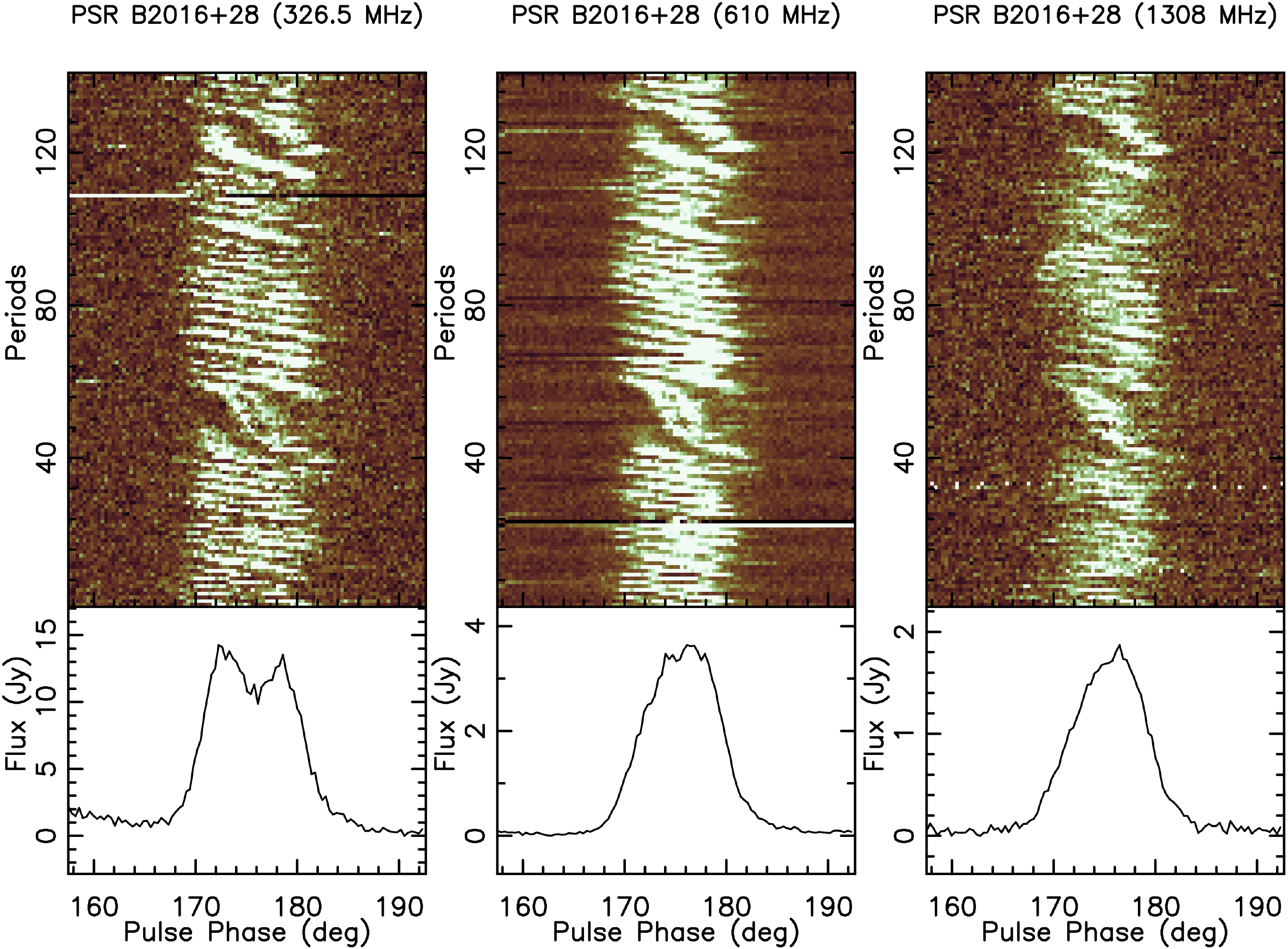}\centering
\caption{Single pulse sequences for PSR B2016$+$28.}
\label{A:2016_sp}
\end{figure*}

\begin{figure*}
\includegraphics[width=0.60\textwidth]{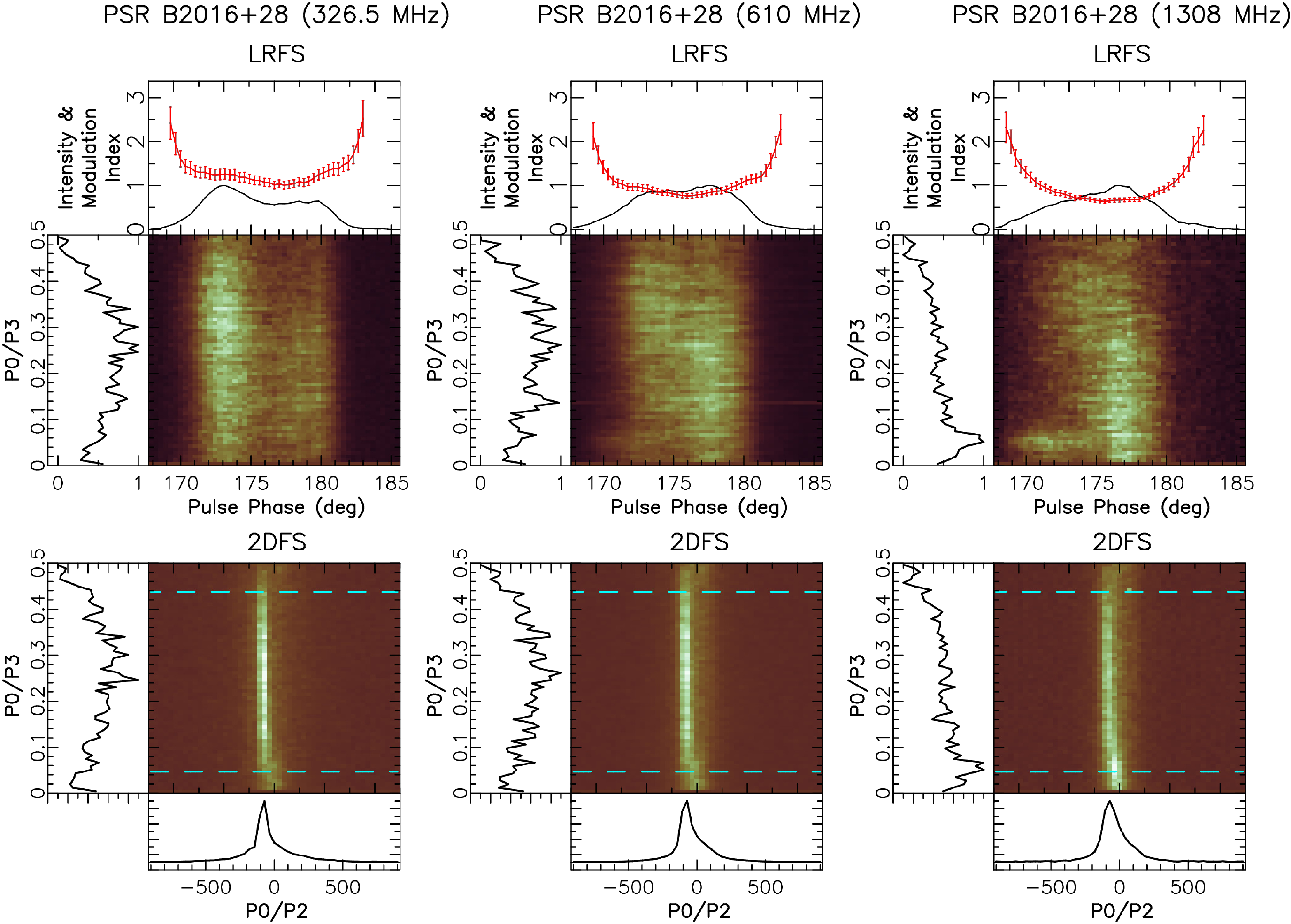}\centering
\caption{LRFS and 2DFS plots for PSR B2016$+$28. The 1308 MHz shows a clear low frequency feature in the leading component in the LRFS plot, which is also present as a weak feature at other frequencies.}
\label{A:2016_2dfs}
\end{figure*}

\begin{figure*}
\includegraphics[width=0.60\textwidth]{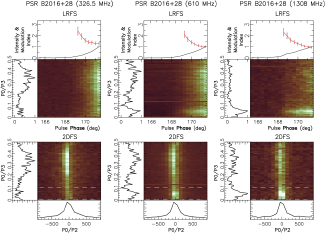}\centering
\caption{LRFS and 2DFS plots for leading component for PSR B2016$+$28. The low frequency feature is clearly seen at 1308 MHz, but is weaker at 326.5 and 610 MHz.}
\label{A:2016_2dfs_comp1}
\end{figure*}
\noindent

\begin{figure*}
\includegraphics[width=0.60\textwidth]{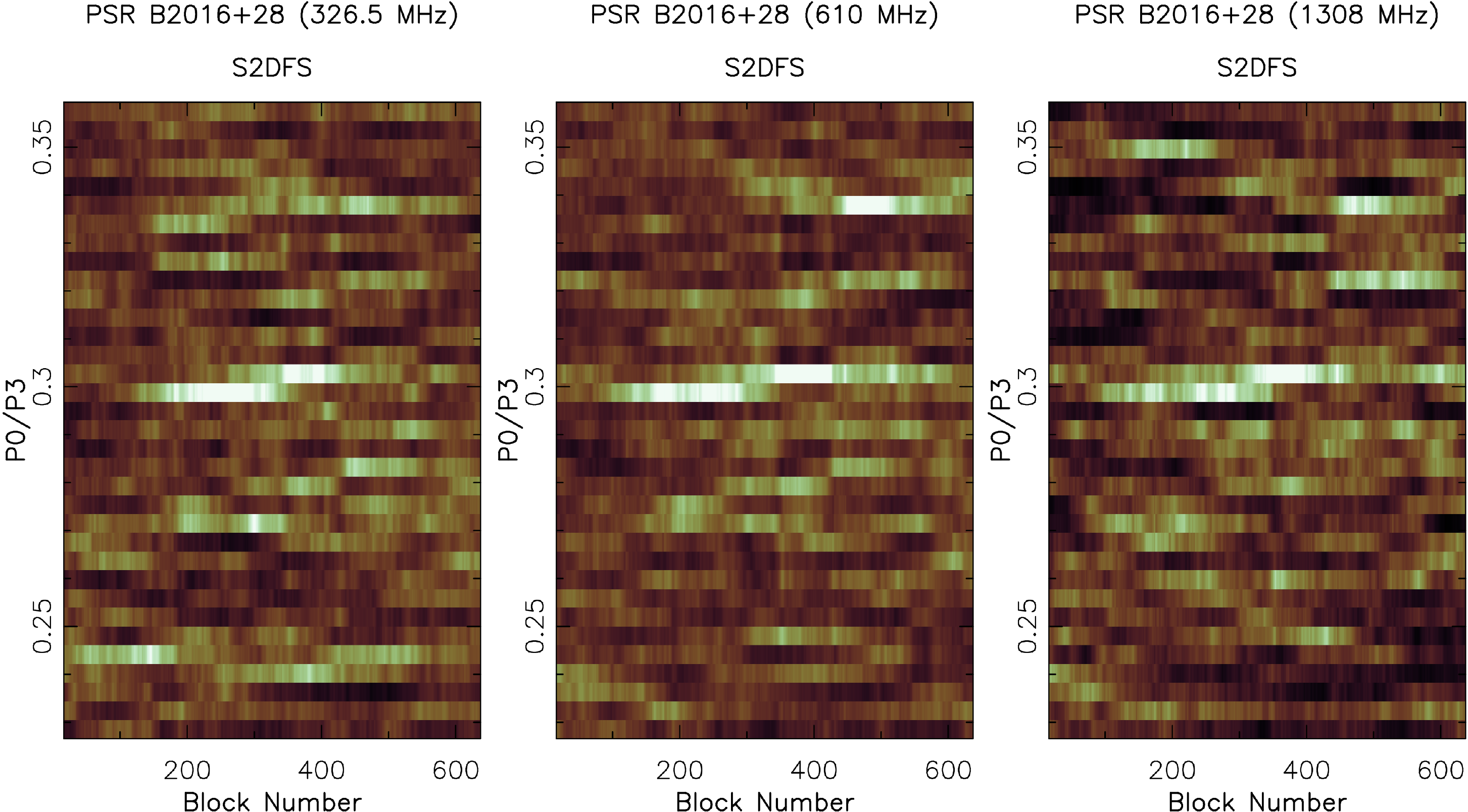}\centering
\caption{S2DFS of PSR B2016$+$28.}
\label{A:2016_s2dfs}
\end{figure*}

\begin{figure*}
\includegraphics[width=0.60\textwidth]{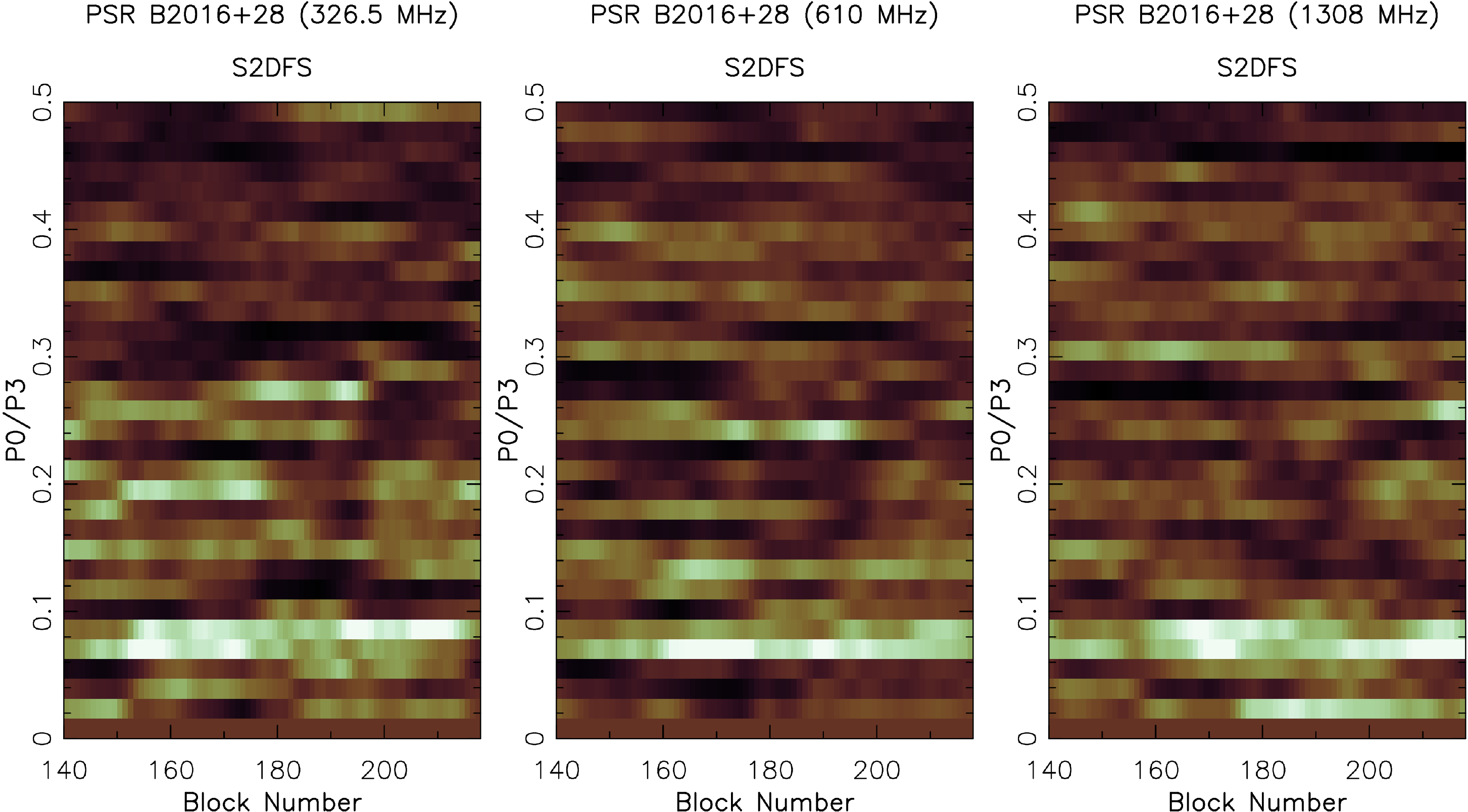}\centering
\caption{S2DFS of leading component of PSR B2016$+$28.}
\label{A:2016_s2dfs_1}
\end{figure*}

\begin{figure*}[!tbp]
  \centering
  \begin{minipage}[c][][b]{0.45\textwidth}
    \includegraphics[width=\textwidth]{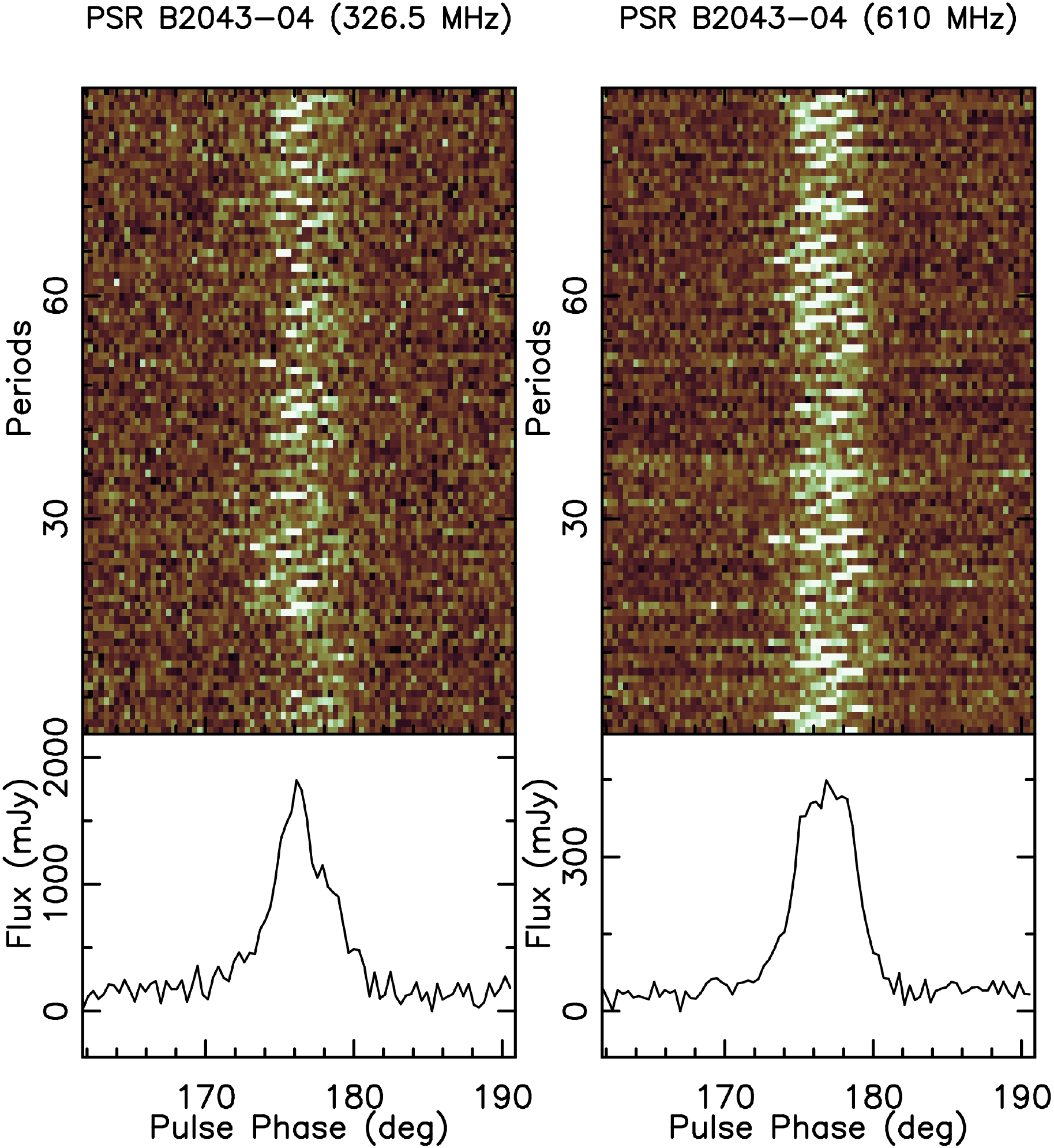}\centering
    \caption{Single pulse sequence for PSR B2043$-$04. The single pulse SNR is low at 1308 MHz observations.}
    \label{A:2043_sp}

  \end{minipage}
  \hspace{0.5cm}
  \begin{minipage}[c][][b]{0.45\textwidth}
   \includegraphics[width=\textwidth]{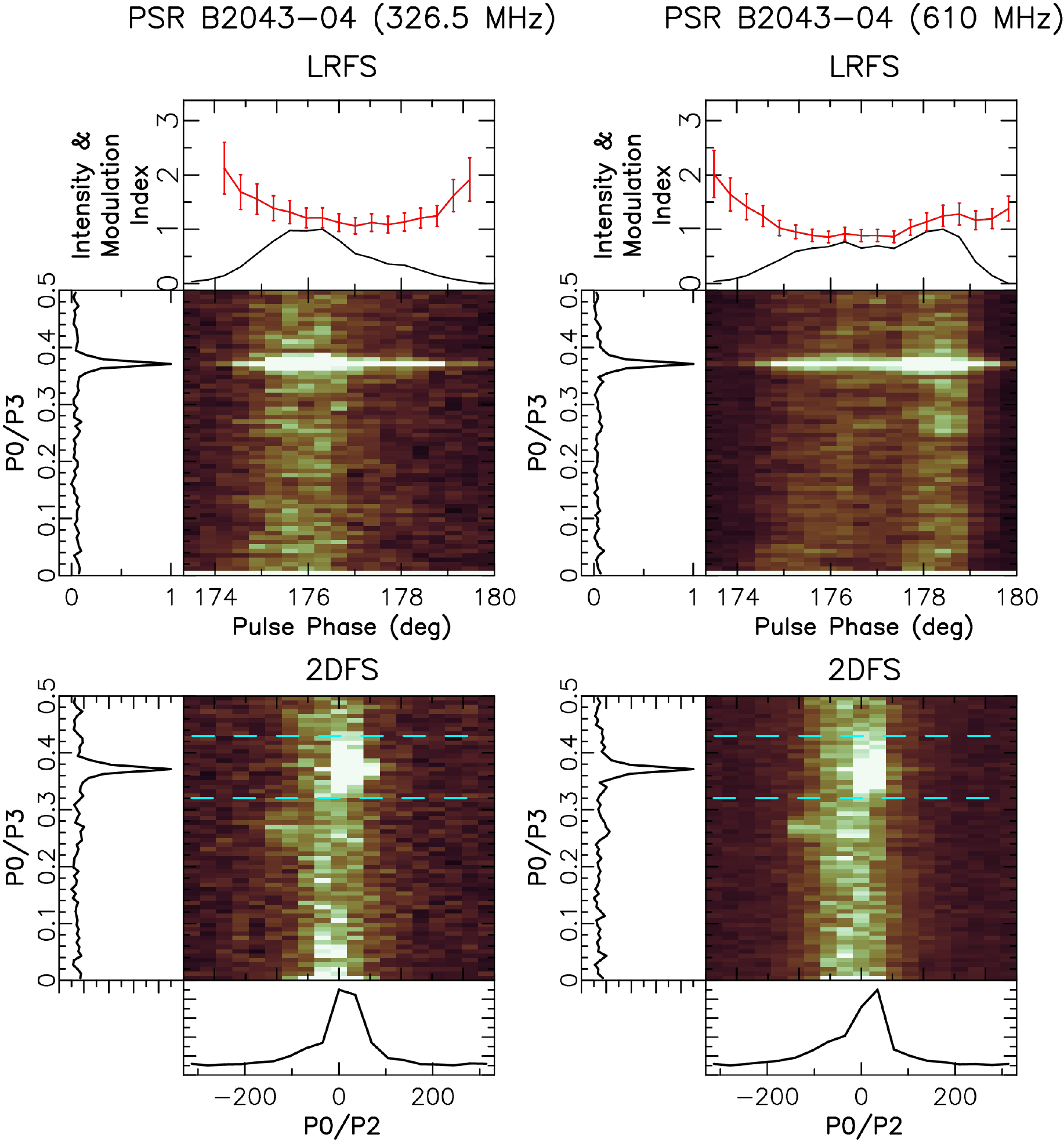}\centering
\caption{LRFS and 2DFS plots for PSR B2043$-$04. Bright narrow features can be seen at 326.5 MHz and 610 MHz. The SNR at 1308 MHz was low, but a very weak detection can be seen in LRFS at 1308 MHz. In addition, there is a very weak feature at about 0.3 cpp at both 610 and 326.5 MHz.}
   \label{A:2043_2dfs}        
  \end{minipage}
   
\end{figure*}

\begin{figure*}[!tbp]
  \centering
  \begin{minipage}[c][][b]{0.45\textwidth}
    \includegraphics[width=\textwidth]{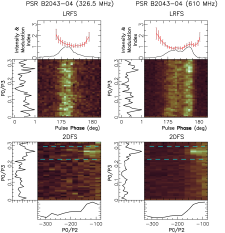}\centering
    \caption{LRFS and 2DFS plots zoomed  around 0.3 cpp for PSR B2043$-$04. A common feature can be seen at both 326.5 and 610.0 MHz.}
    \label{A:2043_2dfs_comp1}

  \end{minipage}
  \hspace{0.5cm}
  \begin{minipage}[c][][b]{0.45\textwidth}
   \includegraphics[width=\textwidth]{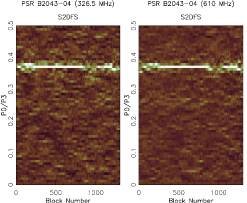}\centering
\caption{S2DFS of PSR B2043$-$04.}
   \label{A:2043_s2dfs}        
  \end{minipage}
   
\end{figure*}

\begin{figure*}
\includegraphics[width=0.60\textwidth]{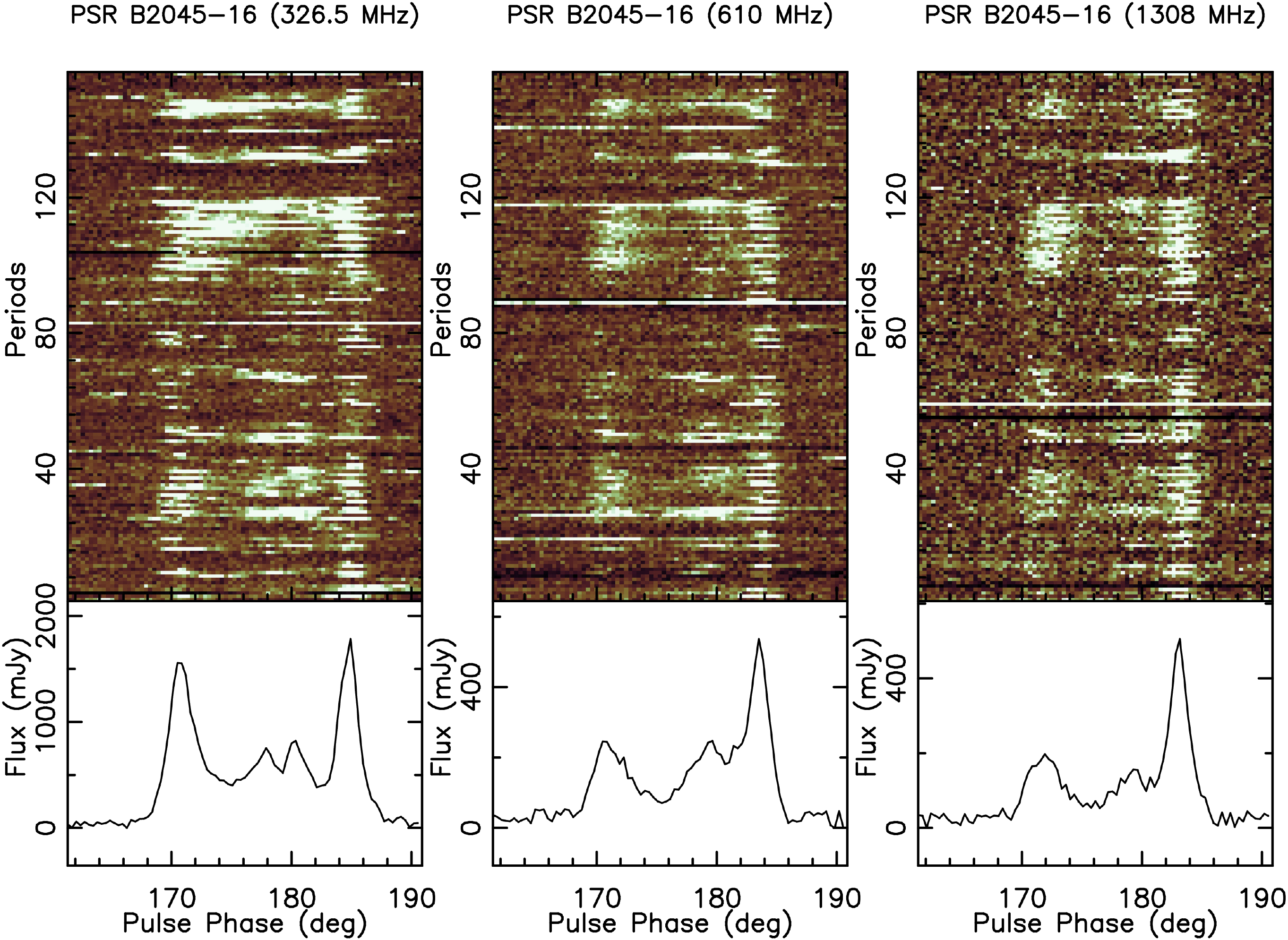}\centering
\caption{Single pulse sequences for PSR B2045$-$16 showing the simultaneous nulla and correlated subpulse behaviour.}
\label{A:2045_sp}
\end{figure*}

\begin{figure*}
\includegraphics[width=0.60\textwidth]{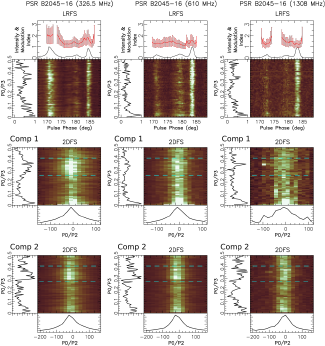}\centering
\caption{LRFS and 2DFS plots for PSR B2045$-$16.}
\label{A:2045_2dfs}
\end{figure*}

\begin{figure*}
\includegraphics[width=0.60\textwidth]{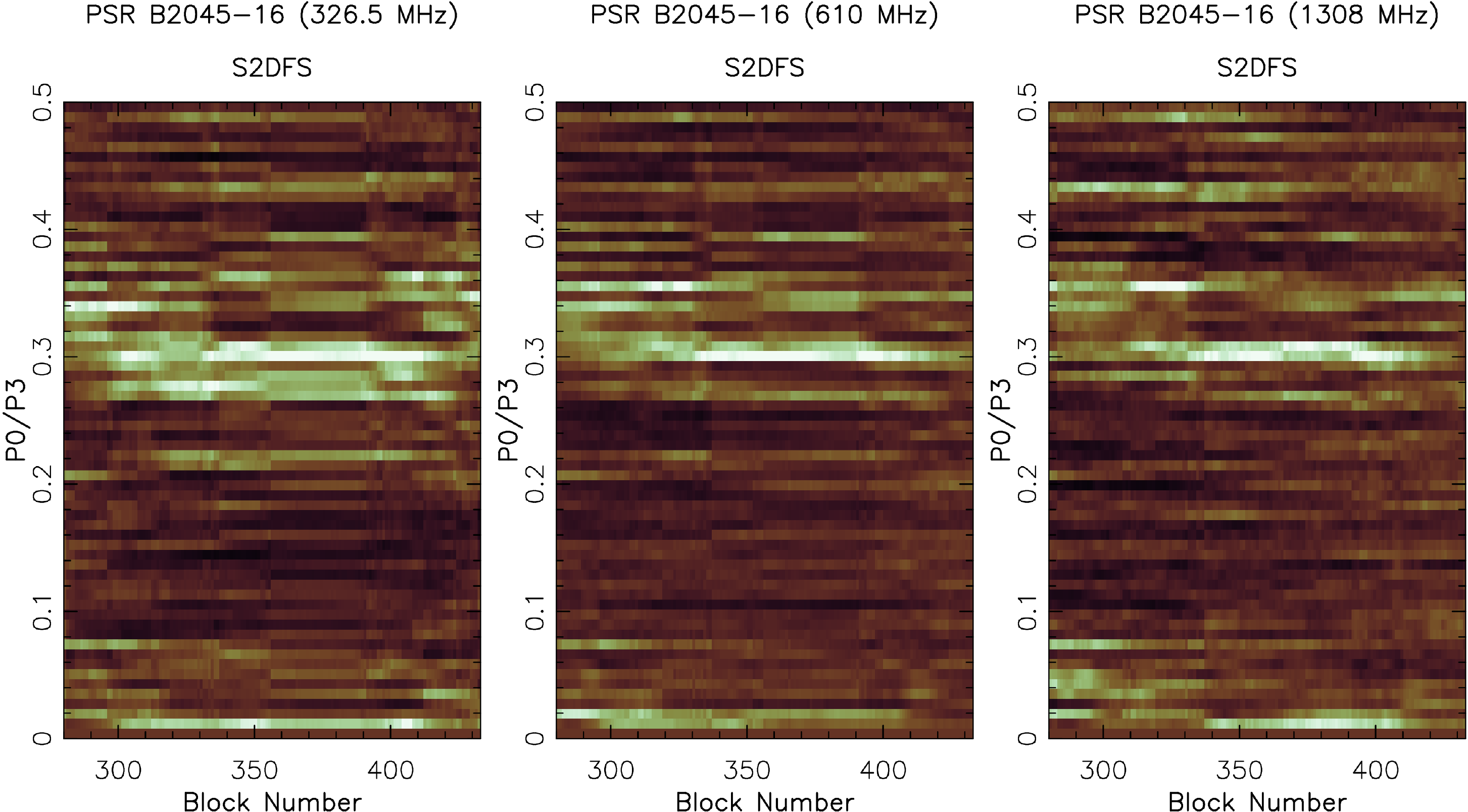}\centering
\caption{S2DFS of PSR B2045$-$16.}
\label{A:2045_s2dfs}
\end{figure*}

\begin{figure*}
\includegraphics[width=0.60\textwidth]{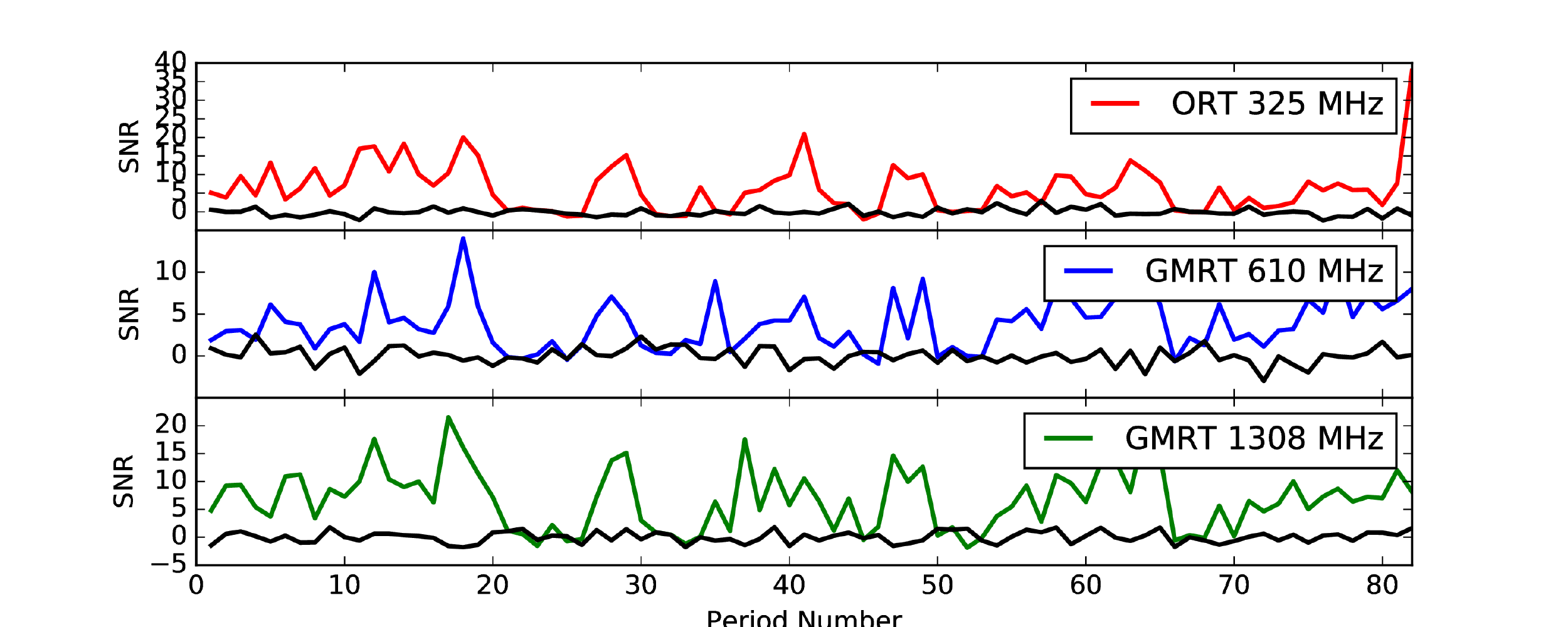}\centering
\caption{On-pulse energy sequence of PSR B2045$-$16.}
\label{A:2045_ep}
\end{figure*}

\end{appendix} 
\end{document}